\documentclass[useAMS,usenatbib]{mn2e}
 
\def\chandra{{\it Chandra}}

\def\Msolar{\hbox{$\rm\thinspace M_{\odot}$}}

\usepackage[dvips]{graphicx}
\usepackage{graphics}
\usepackage{lscape}
\usepackage{amssymb}
\usepackage{mathrsfs}
\input{epsf}

\title[X-ray emission of LBGs]
{The X-ray emission of Lyman break galaxies}
\author[E. S. Laird et al.]
{E. S. Laird,$^{1,2}$\thanks{E-mail:~eslaird@ucolick.org}
K. Nandra,$^{1}$ A. Hobbs,$^{1}$  C. C. Steidel$^{3}$\\ 
$^{1}$Astrophysics Group, Imperial College London, Blackett Laboratory, Prince Consort Road, London SW7 2AZ, UK\\
$^{2}$UCO/Lick Observatory, University of California, Santa Cruz, 1156 High Street, Santa Cruz, CA 95064, USA\\
$^{3}$California Institute of Technology, MS 105-24, Pasadena, CA 91125, USA\\}
                               
\begin{document}

\date{Accepted 0000 December 00. Received 0000 December 00; in original form 0000 October 00}

\pagerange{\pageref{firstpage}--\pageref{lastpage}} \pubyear{2006}

\maketitle

\label{firstpage}
\begin{abstract} 
We present an analysis of the X-ray emission of a large sample of $z\sim3$ Lyman break galaxies (LBGs), based on {\it Chandra}/ACIS  observations of several LBG survey fields. A total of twenty-four LBGs are directly detected in the X-ray, approximately doubling the number of known detections. Thirteen of the LBGs have optical spectroscopic signatures of AGN activity, but almost all the other X-ray detections are also likely to host an accreting black hole based on their X-ray properties. The AGN exhibit a wide range in X-ray luminosity, from weak Seyferts to bright QSOs.  Optical spectroscopy identified approximately 1/3 of the X-ray detected sources as broad line QSOs, 1/3 as narrow line AGN and 1/3 as normal star forming LBGs. The fraction of X-ray detected LBGs is 3 per cent, much lower than has been found for submm selected galaxies. Two galaxies have X-ray luminosities, spectra and f$_{\mathrm {X}}/$f$_{\mathrm {opt}}$ values that are consistent with emission from star formation processes and are identified as candidate X-ray bright, pure starburst galaxies at $z\sim3$. If powered solely by star formation the sources would have SFRs of 300--500~\Msolar~yr$^{-1}$. X-ray spectral analysis of the LBGs shows a mean photon index of  $\Gamma=1.96$, similar to local AGN. There is evidence for absorption in at least 40 per cent of the objects. Significantly more absorption is evident in the narrow line AGN,  which is consistent with AGN unification schemes. After correction for absorption  the narrow and broad line objects show the same average luminosity. X-ray detected LBGs spectroscopically classified as normal galaxies, however, are less luminous in both soft and hard X-ray bands indicating that the host galaxy is outshining any optical AGN signature. Turning to the X-ray emission from LBGs without direct detections, stacking the X-ray flux in the two deepest {\it Chandra} fields under consideration (the HDF-N and GWS) produced significant detections in each, although the GWS result was marginal. The detection in the HDF-N  gives an X-ray derived SFR of $42.4\pm7.8$M$_{\rm \odot}$~yr$^{-1}$ per LBG and, by comparing with the UV SFR, the implied  UV extinction correction is $4.1 \pm 0.8$. The LBG sample was split into three bins based on UV magnitude to examine the correlation between UV and X-ray emission: for the limited statistics available there was no evidence of any correlation. 
\end{abstract}


%%%%%%%%%%%%%%%%%%%%%%%%%%%%%%%%%%%%%%%%%%%%%%%%%%
\section{Introduction}

Observations of distant galaxies are an essential component in our understanding of the development of the Universe. Most high-redshift galaxies have been identified in the UV via the Lyman break technique (e.g. \citealt{steidel96,lowenthal97,steidel03}).  There is therefore considerable interest in constraining properties of LBGs such as the star-formation rate (SFR), dust content, stellar mass and metallicity. Another effective way of identifying high-$z$ galaxies is by observations at sub-mm wavelengths with SCUBA on the James Clerk Maxwell Telescope (e.g. \citealt{ivison02,smail02,chapman03}). The relation between LBGs and SCUBA galaxies is still not entirely clear but studies have shown that up to 50 per cent of SCUBA galaxies have similar rest-frame UV colours as LBGs \citep{reddy06,steidel04}. LBGs and SCUBA galaxies may simply form a continuous distribution in SFR and dust content, with SCUBA galaxies occupying the upper end of the distribution \citep{reddy06,reddy05,adelberger00}. Indeed a recent study of IR luminous LBGs showed that while LBGs have a diverse range of IR properties, at least some LBGs have the same IR properties as SCUBA galaxies and are likely to be the progenitors of today's massive giant ellipticals \citep{huang05}. 

The proposed starburst-AGN connection that arises out of merger-driven galaxy formation scenarios (e.g. \citealt{sanders96,hopkins05}),  and which is a natural explanation for the observed $M_{BH}-\sigma$ relation between the mass of dormant black holes and galaxy bulges  in local galaxies  (e.g. \citealt{magorrian98,gebhardt00,ferrarese00}), also leads to an interest in the AGN content of high-$z$ star forming  galaxies. The incidence of AGN in SCUBA sources  has attracted considerable attention (e.g. \citealt{fabian00,barger01}).  Most recently, \citet{alexander05} have found an extremely high rate of X-ray detection of radio-identified SCUBA galaxies, inferring that at least 40 per cent, and perhaps as many as 75 per cent of submm-selected galaxies contain an AGN.  Much less work has been done on the properties of AGN in the LBG population. To date most work including X-ray data has focussed on SFRs and extinction corrections of LBGs without AGN (e.g. \citealt{brandt01,N02,lehmer05}). \citet{steidel02} and \citet{hunt04}
analysed rest-frame UV spectra of LBGs to identify AGN based on emission lines and to calculate the faint end of the AGN luminosity function at high-$z$.  They found that 3 per cent of LBGs harbour AGN, which contribute 8 per cent to the integrated 1500~\AA~UV luminosity at $z=3$. \citet*{nandra05a} used LBG selection in two fields (the HDF-N and GWS) as a method of efficiently identifying the redshift of AGN in X-ray surveys. The results were used to estimate the space density of moderate luminosity AGN at $z=3$, finding them to be ten times more common than high luminosity QSOs.  
  
In this paper the X-ray emission of $z\sim3$ LBGs in six of the survey fields presented by \citet{steidel03}  are analyzed. This work extends the analysis of LBGs in the 1~Ms CDF-N by Nandra et al. (2002; hereafter N02)  and \citet{nandra05a} by using more LBG fields and including X-ray spectral analysis of the LBGs. Following on from the stacking analysis of LBGs by \citet{brandt01}, N02 and \citet{lehmer05}, the X-ray emission from non-AGN dominated LBGs in the two deepest fields is also analyzed.  

Throughout, a standard, flat $\Lambda$CDM cosmology with $\Omega_\Lambda = 0.7$ and H$_0 = 70$~km~s$^{-1}$~Mpc$^{-1}$ is assumed.

%%%%%%%%%%%%%%%%%%%%%%%%%%%%%%%%%%%%%%%%%%%%%%%%%%
\section{Data and source sample}

%%%%%%%%%%%%%%%%%%%%%%%%%%%%%%%%%%%%%%%%%%%%%%%%%%
\subsection{Optical LBG data}

Our sample of $z$$\sim$3 LBGs is culled from the \citet{steidel03} survey fields that have archived \textit{Chandra} ACIS imaging data. These fields are the \textit{Hubble} Deep Field-North (HDF-N), Groth-Westphal Strip (GWS), Q1422+2309, SSA22 (SSA22a and SSA22b), B20902+34 and 3C 324. 

The areas of overlap between the LBG fields and the \textit{Chandra} data, along with the number of LBGs in the overlap areas are shown in Table~\ref{table1}.  The LBG data for fields B20902+34 and 3C 324, for which the targets of both were radio galaxies, covers a small area and therefore contain relatively few LBG candidates. Q1422+2309, which was centred on a $z=3.620$ gravitationally lensed QSO, has the deepest data of all the \citet{steidel03} survey fields.

In the HDF-N region, imaging for LBG selection and spectroscopy has recently been extended  by the Steidel team from the original survey area to now cover the larger GOODS-N region. The catalogue of HDF-N LBGs used in this paper consists of all the LBGS from \citet{steidel03} plus the LBGs from the extended survey area. The latest imaging and  photometry is of better quality than the original data and is therefore used for any LBG candidates in the original list that are also covered by the new data. 

Of the LBGs considered in this work 43 per cent have confirmed spectroscopic identifications as $z>2$ galaxies. Over the whole LBG survey (containing more
than 2300 objects) the redshift distribution of candidates is $z=2.96
\pm 0.29$ and the fraction of low-redshift interlopers is only
4 per cent. Given the tightly peaked redshift distribution and the low interloper fraction we are confident
that the remainder of our sample without spectroscopic confirmation are indeed high-redshift star-forming galaxies.

%%%%%%%%%%%%%%%%%%%%%%%%%
\begin{table*}
\begin{minipage}{180mm}
\begin{center}
\caption{Fields included in analysis.
Col.(1): LBG field name;
Col.(2): Type of \textit{Chandra} observation;
Cols.(3, 4): Nominal Right Ascension and Declination of \textit{Chandra} pointings; 
Col.(5): Galactic column density;
Col.(6): Exposure time after GTI, background filtering etc;
Col.(7): Area of overlap between LBG survey and \textit{Chandra} observations; 
Col.(8) Number of LBGs in combined X-ray--LBG survey area; 
Col.(9) Number of spectroscopically identified LBGs with $z$$>$2.
}
\label{table1}
\begin{tabular}{@{}lcccccccc@{}}
\hline
Field & \textit{Chandra} & RA & Dec & N$_{\rmn{H}}$ & Filtered Exposure & Survey area & No. of  &  No. of LBGs\\
      & observation & (J2000) & (J2000) & ($10^{20}\rmn{cm}^{-2}$) & (ks)           & (arcminutes$^2$)  & LBGs &  with $z_{\rmn{spec}}$  \\                
(1)   & (2) & (3) & (4)               & (5)         & (6)            & (7) & (8) & (9)\\ 
\hline
HDF-N    & ACIS-I & 12:36:47.59 & +62:14:08.06 &1.6$^a$ &1862.9 & 149.1 &  295 & 93  \\
GWS      & ACIS-I & 14:17:43.04 & +52:28:25.20 &1.3$^b$& 190.6  & 239.0 &  334 & 200 \\
Q1422+2309& ACIS-S & 14:24:35.61 & +22:55:43.76 & 2.7$^b$&28.4   & 78.1  &  292 & 103 \\
SSA22    & ACIS-S & 22:17:28.24 & +00:15:09.59 &4.7$^b$ &77.8   & 105.5 &  184 & 86  \\
B20902+34 & ACIS-S & 09:05:33.28 & +34:09:07.83 &2.3$^b$& 9.78   & 41.8  &  76  & 38  \\
3C 324   & ACIS-S & 15:49:46.41 & +21:25:19.58 & 4.3$^b$ &38.5   & 36.9  &  45  & 9   \\
\hline
\end{tabular}
\end{center}
$^a$ \citealt{stark92}\\
$^b$ \citealt{dickey90}
\end{minipage}
\end{table*}
%%%%%%%%%%%%%%%%%%%%%%%%%

%%%%%%%%%%%%%%%%%%%%%%%%%%%%%%%%%%%%%%%%%%%%%%%%%%
\subsection{X-ray data and reduction}

%%%%%%%%%%%%%%%%%%%%%%%%%%%%%%%%%%%%%%
\begin{figure}
\begin{center}
\includegraphics[width=85mm]{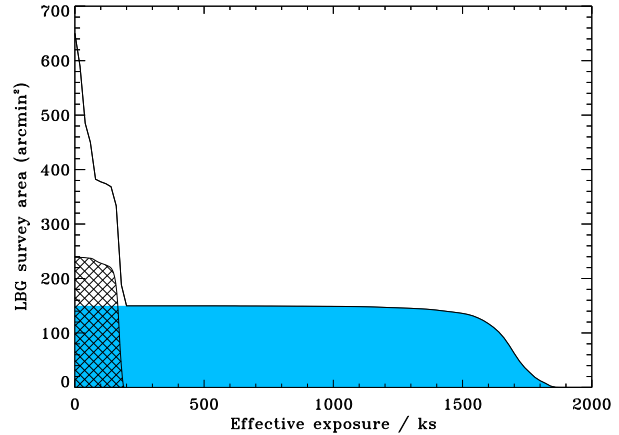}
\caption{Survey area versus the minimum effective exposure in the soft band exposure maps for all of the LBG fields included in this work. The hatched and shaded sections show the survey area versus minimum effective exposure for the LBG regions in the GWS and HDF-N fields, respectively, }
\label{figure1}
\end{center}
\end{figure}
%%%%%%%%%%%%%%%%%%%%%%%%%%%%%%%%%%%%%%%

The majority of the results in this paper come from the deep X-ray observations of the HDF-N and GWS, with the remaining fields acting as a supplementary data set. The \textit{Chandra} ACIS-I observation of the HDF-N (the \textit{Chandra} Deep Field-North, CDF-N), with a total exposure time of approximately 2~Ms, is the deepest X-ray observation taken to date.  A full description of the data and point source analysis was presented by Alexander et al. (2003; hereafter A03). For this work we used the raw data available from the public archives. Details of the data reduction can be found in \citet{L05}.
The 200~ks \textit{Chandra} ACIS-I survey of the GWS is currently the third deepest blank-field , extragalactic \textit{Chandra}  survey field, after the HDF-N and \textit{Chandra} Deep Field-South (CDF-S). A description of the data and reduction, including a point source list, is detailed in \cite{nandra05b}. 

The remaining LBG fields -- Q1422+2309, SSA22, B20902+34 and 3C 324 -- were each observed once with the \textit{Chandra} ACIS-S instrument at prime focus, with exposure times ranging from $\sim$10~ks to $\sim$80~ks.  Details of the observations are summarised in Tables~\ref{table1} and \ref{table2}. The data were reduced using the \textit{Chandra} X-ray Center
(CXC) \textit{Chandra} Interactive Analysis of Observations ({\small CIAO}) software, version 3.1, and  the \textit{Chandra} calibration database (CALDB) version 2.27.  The reduction procedure was very similar to that employed with the HDF-N and GWS data, but more straight forward due to each target being observed just once. Images and event files were created for the same four analysis bands as used in the HDF-N and GWS analyses: 0.5--2~keV (soft band), 0.5--7~keV (full), 2--7~keV (hard) and 4--7~keV (ultra-hard). Exposure maps for the LBG survey areas of the fields were made using {\small MERGE\_ALL}  with the same representative energies that were used for the HDF-N and GWS data -- namely 1~keV, 2.5~keV, 4~keV and 5.5~keV for the soft, full, hard and ultra-hard bands respectively. Filtering to exclude periods of unstable background using {\small ANALYZE\_LTCRV} removed 0.4~ks,  1.1~ks, 0.1~ks and  4.2~ks from Q1422+2309, SSA22, B20902+34 and 3C 324, respectively. 

The final filtered exposure times and aim points of the \textit{Chandra} data are shown in Table~\ref{table1}. Figure~\ref{figure1} shows the cumulative survey solid angle as a function of effective exposure over the entire \textit{Chandra}/LBG survey area.

%%%%%%%%%%%%%%%%%%%%%%%%%%%%%%%%%%%%%%
\begin{table}
\begin{center}
\caption{Table of ACIS-S observations of LBG fields. Each field was observed only once. Col.(1): LBG field name; Col.(2): \textit{Chandra} observation identification number; Col.(3): Date of beginning of observation; Col.(4): Nominal roll angle of satellite (degrees East of North).}
\label{table2}
\begin{tabular}{@{} lcccc@{}}
\hline
Field &Obs. ID & Date &  Roll  \\
Name        & & (UT)  &Angle \\  
(1)      & (2)     & (3)           & (4)       \\         
 \hline
Q1422+2309 &367 &2000-06-01&218.6  \\
SSA22 &1694 &2001-07-10& 120.3    \\
B20902+34 &1596 &2000-10-26& 71.0 \\
3C 324 & 326 &2000-06-25& 221.1  \\
\hline
\end{tabular}
\end{center}

\end{table}
%%%%%%%%%%%%%%%%%%%%%%%%%%%%%%%%%%%%%%

%%%%%%%%%%%%%%%%%%%%%%%%%%%%%%%%%%%%%%%%%%%%%%%%%%
\subsection{Source detection, photometry and matching}

Source detection and photometry was carried out using our own procedure which is described  in \citet{nandra05b}. As this work is only interested in the X-ray counterparts to the LBGs we set the detection threshold to the low value of $10^{-4}$. This is a less stringent value than one which would be acceptable for analysis of the general X-ray source population of our fields. Source detection was carried out in the four bands described in \S2.2 and a band merged catalogue produced. A detection with Poisson probability less than $10^{-4}$ is required in at least one band. 
For all the sources in the band merged catalogue accurate source photometry is determined
using 90 per cent EER apertures in the full, hard and ultra-hard bands and 95 per cent EER apertures  in the soft band. Upper and lower confidence regions for the background subtracted counts are calculated for 1$\sigma$ statistical errors (according to equations 7 and 14 of \citealt{gehrels86}). If the Poisson probability in a band is greater than or equal to $1.3\times10^{-3}$ (equivalent to a Gaussian 3$\sigma$ detection) then we calculate the upper limit to the counts based on that same probability. 

The PSFs were calculated at the representative energies of each of the four bands, according to the procedure described in \citet{nandra05b}.  In the HDF-N we use exposure weighted averages of the individual PSFs (see \citealt{L05}). To convert count rates to fluxes we assume a power-law source spectrum with photon index $\Gamma$$=$1.4 and Galactic column density (Table~\ref{table1}). Fluxes in the full, hard and ultra-hard bands were extrapolated to the standard upper limit of 10~keV. The effects of the \textit{Chandra} ACIS quantum efficiency (QE) degradation \citep{marshall04}) have been taken into account in all fluxes quoted in this paper.

The X-ray catalogues were cross-correlated to the LBG candidates using the following procedure. For each field we first matched the catalogues using a search radius of 2.0 arcsec to identify possible counterparts. Any astrometric offsets between the X-ray and LBG reference frames were then identified and the LBG positions corrected accordingly. Astrometric offsets were present in all the fields, with the necessary shifts varying from 0.16 arcsec (HDF-N) to 2.06 arcsec (GWS). After correcting for the overall offsets the \textit{Chandra} and LBGs catalogues were re-matched using a radius of 1.5 arcsec. 

The false detection and match rate was assessed by creating false LBG catalogues for each field. Using 200 000 random `galaxy' positions and matching to the X-ray catalogues using a radius of 1.5 arcsec we find that the probability of detecting an X-ray source at the position of a known LBG at random is 0.004, 0.001, 0.002 and 0.001 for the HDF-N, GWS, SSA22 and Q1422+2309 fields, respectively. Given the number of LBGs and X-ray sources in the fields this corresponds to an expected 0.92, 0.42, 0.34 and 0.33 false matches in the HDF-N, GWS, SSA22 and Q1422+2309 fields, respectively.

%%%%%%%%%%%%%%%%%%%%%%%%%%%%%%%%%%%%%%%%%%%%%%%%%%
\section{Basic properties of X-ray detected LBGs}

A total of 24 LBGs in the six fields were found to have significant X-ray emission -- ten in the HDF-N, six in the GWS, three in Q1422+2309 and five in SSA22a. We did not find any X-ray counterparts to the LBGs in fields SSA22b, B20902+34 and 3C 324.  Twenty of the LBGs have confirmed redshifts with $2.211<z<3.630$. 
All of the LBGs that were identified as QSO or AGN by their rest-frame UV spectra in the 
HDF-N and SSA22 fields have been detected in the \textit{Chandra} data. 
Table~\ref{table3} presents the basic X-ray and optical properties of the sources, including coordinates, $\mathscr{R}$ magnitude, redshift, X-ray fluxes and hardness ratio (HR).

In the HDF-N, only one of the five least luminous X-ray detected LBGs, CXO~J123622.5+621306 
(=HDF-C14), has already been reported as an X-ray source in 
the 2~Ms CDF-N catalogue of A03 (source 133). An
IR--optical counterpart  was detected in bands HK' through U by \citet{barger03} and a spectrum
was taken but no identification or spectroscopic redshift could be
obtained. \citet{steidel03} classified this LBG as a normal galaxy at $z$=2.981.
The remaining four LBGs, CXO~J123618.4+621139 (=HDF-D7),
CXO~J1236704.2+624446 (=HDF-oMD49), CXO J123645.0+621653 (=HDF-M35)  and 
CXO J123651.5+621041 (=HDF-M9), were not in the A03 catalogue which covered the whole 
of the CDF-N area and had a more conservative detection threshold. As discussed in \S2.3, less stringent detection thresholds, which result in lower flux limits, can be used when searching only for the counterparts to limited numbers of known sources as is the case in this work.

The 0.5--2~keV flux of the LBGs varies over two orders of magnitude from 
$<3\times10^{-17}$~erg~cm$^{-2}$~s$^{-1}$ to  $7.4\times10^{-15}$~erg~cm$^{-2}$~s$^{-1}$. 
The $\mathscr{R}$ band flux (corresponding to rest frame $\sim$1800 \AA~emission) covers as large a range, from the bright $\mathscr{R}$=20.48 QSO, HDF-MD39, to very faint $\mathscr{R}$=25.27 galaxy, SSA22a-M14, associated with the Ly$\alpha$ nebula.

%%%%%%%%%%%%%%%%%%%%%%%%%%%%%%%%%%%%%%
\begin{figure}
\begin{center}
\includegraphics[width=85mm]{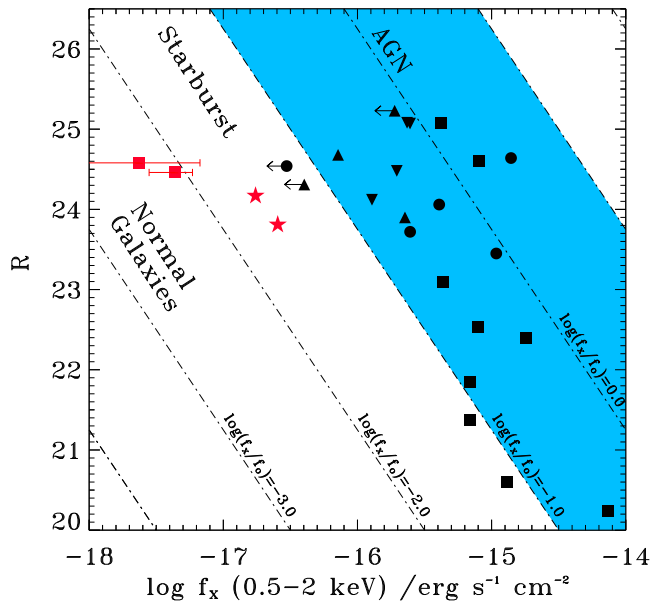}
\caption{Soft band X-ray flux vs. R magnitude for the 24
X-ray detected LBGs. Black squares and circles denote optically identified broad line QSOs and optically identified narrow line AGN, respectively. Black triangles denote those LBGs with optical galaxy classifications (GAL). Black inverted triangles denotes LBGs without optical spectra. The red stars identify  HDF-M9 and HDF-M35 which are GAL LBGs with X-ray fluxes and spectra that do not clearly indicate the presence of an AGN and are consistent with emission from star formation.
The red squares show the soft band stacking results for the undetected LBGs in the CDF-N and GWS fields. Arrows signify the 3$\sigma$ upper limit of sources
not detected in the soft band. 
$\mathscr{R}$ band magnitudes have been converted to Kron-Cousins R.
The diagonal lines indicate constant X-ray to optical flux ratios, as indicated.}
\label{figure2}
\end{center}
\end{figure} 
%%%%%%%%%%%%%%%%%%%%%%%%%%%%%%%%%%%%%%%

Figure \ref{figure2}  shows $R$ magnitude versus 0.5--2~keV flux
for the detected LBGs. $\mathscr{R}$ magnitudes  were converted to 
Kron-Cousins $R$ using the conversion in 
\citet{steidel93}. Lines of constant X-ray to optical flux are also shown,  according to the relation from \citet{horn01}:
\begin{equation}
\rmn{log} \left( \frac{f_{\rmn{X}}}{f_{\rmn{R}}} \right) =
\rmn{log} \; f_{\rmn{X}} + 5.50 + \frac{\rmn{R}}{2.5}.
\end{equation}
The ratio of X-ray to optical flux is a
commonly used technique for determining the nature of
X-ray sources (e.g. \citealt{macc88}; \citealt{stocke91};
\citealt{schmidt98}). Over several decades in flux classic AGN,
including luminous narrow line AGN (NLAGN) as well as broad line AGN
(BLAGN), exhibit X-ray to optical flux ratios of $-1 <
\rmn{log\;(f_{X}/f_{opt}) < +1}$ in both soft and hard bands
(e.g. \citealt{schmidt98}; \citealt{akiyama00}; \citealt{lehmann01};
\citealt{alexander01}). At lower X-ray to optical flux ratios, mainly
in the soft band at $f_{0.5-2 \rmn{keV}} \lesssim 10^{15}$~erg
s$^{-1}$ cm$^{-2}$, starburst galaxies, normal galaxies, low
luminosity and heavily absorbed AGN emerge (e.g. \citealt{giacconi01};
\citealt{alexander01}; \citealt{barger02}; \citealt{barger03}). 
Starburst galaxies and AGN populate the $-2<\rmn{log\;(f_{X}/f_{opt}) < -1}$ range (e.g. \citealt{bauer02};
\citealt{alexander02}) while normal galaxies which have very weak
X-ray emission dominate at $\rmn{log\;(f_{X}/f_{opt}) <-2}$
\citep{horn03}. 

The X-ray to optical flux of the majority of the LBGs falls within the region populated by classic, luminous AGN (shaded region). Two of the QSOs and one AGN are sub-luminous in X-ray compared to their optical emission for typical active galaxies. Half of the LBGs identified as normal galaxies based on their UV spectra have X-ray to optical flux ratios typical of NLAGN and BLAGN.  Two LBGs, HDF-M9 and HDF-M35, exhibit ratios typical of starburst galaxies (see \S6 for further discussion).
As a cautionary note, the location of the LBGs in Figure \ref{figure2} could be affected by K-corrections which have been shown to have a redshift dependent effect on $\rmn{log\;(f_{X}/f_{opt})}$ values for Seyfert galaxies \citep{peterson06}.

\section{X-ray spectral analysis}

\subsection{Hardness Ratios}

To assess the X-ray spectral slope of the LBGs, based on broad band X-ray photometry, the effective photon index ($\Gamma$) was calculated for each LBG using the observed soft-to-hard band ratio. The calculations were made using {\small PIMMS} v3.6 and assuming only Galactic absorption.  The results for sources with meaningful soft-to-hard band ratios (i.e. excluding sources with only full band detections or those with no hard band detections and soft band fluxes close to  the detection limit)  are shown in Table~\ref{table4}. As an alternative measure of the X-ray spectrum, the intrinsic column density required to produce the observed band ratio for an underlying assumed power-law spectrum with $\Gamma=2$  was also calculated  (Table~\ref{table4}).  A $\Gamma\simeq2$ power-law is the canonical spectrum of unobscured AGN (e.g. \citealt{nandra94}; \citealt{george00}) and additionally well represents the average 2--10~keV spectrum of star forming galaxies \citep{ptak99}.

The effective $\Gamma$ of the LBGs covers a wide range from $>$2.2 for SSA22a-D13 (indicating a soft, unabsorbed spectrum)  to $<$0.8 for SSA22a-M14 (representative of a flat or absorbed spectrum). Alternatively, assuming an underlying continuum with a $\Gamma$=2 power-law, the LBGs have intrinsic column densities of  N$_{\rmn {H}}\sim0.6 - 3.0\times10^{23}$ cm$^{-2}$.

%%%%%%%%%%%%%%%%%%%%%%%%%
\begin{table}
\begin{center}
\caption{Effective photon indices ($\Gamma$) and, alternatively, effective intrinsic column densities (N$_{\rmn H}$) calculated from the hard-to-soft band ratios, for sources with meaningful limits. Col.(1): LBG name; Col.(2): Effective $\Gamma$ over 0.5--10 keV  assuming only Galactic absorption. Col.(3): Intrinsic column density calculated for an underlying power-law spectrum with $\Gamma=2$ in units of $10^{23}$~cm$^{-2}$.
}
\label{table4}
\begin{tabular}{@{}lcc@{}}
\hline
LBG  &  $\Gamma$ & N$_{\rmn{H}}$  (for $\Gamma=2$)  \\
 & & (10$^{23}$~cm$^{-2}$) \\
(1)   & (2) & (3)            \\
\hline
HDF-D7   & $<1.0 $ & $>2.7$\\
HDF-C14  & $1.3 ^{+0.3 }_{-0.2 } $& $2.0^{+1.1}_{-0.9}$\\
HDF-oC34 & $1.6 ^{+0.1 }_{-0.1 } $  & $1.3^{+0.5}_{-0.5}$\\
HDF-C10  & $1.6 ^{+0.2 }_{-0.2 } $& $1.0^{+0.7}_{-0.3}$\\
HDF-MD34 & $1.7 ^{+0.1 }_{-0.2 } $ & $0.7^{+0.4}_{-0.3}$\\
HDF-MD12 & $1.2 ^{+0.1 }_{-0.1 } $    & $2.3^{+0.3}_{-0.2}$\\     
HDF-MD39 & $1.6 ^{+0.1 }_{-0.1 } $& $0.6^{+0.1}_{-0.1}$\\
Q1422-C73  & $1.3 ^{+0.4 }_{-0.3 }$ & $2.5^{+2.5}_{-1.8}$\\
SSA22a-D13 & $>2.2 $ & No abs.\\
SSA22a-D12 & $1.3 ^{+0.4 }_{-0.3 } $ & $2.3^{+1.9}_{-1.5}$\\
SSA22a-M14 & $<0.8 $ & $>5.6$ \\
GWS-MD106 & $1.6 ^{+0.2 }_{-0.1 } $ & $0.8^{+0.4}_{-0.4}$ \\
GWS-D54   & $1.4 ^{+0.2 }_{-0.2 } $ & $1.9^{+0.9}_{-0.9}$\\
GWS-M47   & $1.1 ^{+0.3 }_{-0.2 } $ & $3.0^{+1.6}_{-1.5}$\\
GWS-C50   & $1.2 ^{+0.4 }_{-0.3 } $ &  $2.4^{+1.8}_{-1.6}$ \\
\hline
\end{tabular}
\end{center}
\end{table}
%%%%%%%%%%%%%%%%%%%%%%%%%

%%%%%%%%%%%%%%%%%%%%%%%%%%%%%%%%%%%%%%%%%%%%%%%%%%

\subsection{Spectral fitting}

To investigate further the effects of absorption, and see if there is a correlation between X-ray spectral properties and the optical spectral classifications, X-ray spectral analysis was carried out.
{\small CIAO} v3.2.1 and {\small CALDB} v3.0.1 were used for the spectral extraction and {\small XSPEC} v11.3.1 software used for the spectral analysis. The energy range of all of the analysis was restricted to 0.5--7~keV.

For sources in the GWS, SSA22 and Q1422 fields,  the source and background spectra were extracted  using the {\small CIAO} tool {\small PSEXTRACT}. The source spectra were extracted from a circular aperture with a 95 per cent EER calculated at 2.5~keV. Local background regions were manually selected to avoid contamination by nearby X-ray sources. Typically, the background was extracted from three to four large regions surrounding the source.

As a result of the different aim points, roll angles and observing modes of the HDF-N observations the spectra in this field could not be extracted using {\small PSEXTRACT} in one step. Instead, source and local background regions were defined as before and spectra were extracted for each of the individual observations. The spectra were then summed using the standard {\small FTOOLS} \citep{blackburn95} routine {\small MATHPHA}. The instrument response and matrix files were coadded using the {\small FTOOLS} routines {\small ADDARF} and {\small ADDRMF}, weighted by exposure time. The response files for the first three HDF-N observations (IDs 580, 966 and 967) are different to those for the remaining observation IDs as a result of the higher focal plane temperature for these observations. Therefore when combining the individual spectra only the 17 observations taken at $-120$\degr C were included, reducing the total exposure of the spectra by 161.7~ks.

Most of the sources have limited counting statistics ($< 200$ counts) so the spectral analysis was performed using the C-statistic \citep{cash79}, which was specifically developed to extract information from spectra with low numbers of counts.  The sample was limited to sources that had a minimum of 10 total counts in the extracted spectra, resulting in the exclusion of SSA22a-M8, SSA22a-M14, SSA22a-MD14 and Q1422+2309b from the spectral analysis. All of the spectra analysed using the C-statistic were grouped into fixed width bins of 4 channels ($\sim700$~eV), which improves the processing times in {\small XSPEC}.  For those sources with $>100$ counts in their spectra (HDF-oC34, HDF-MD34, HDF-MD12, HDF-MD39 and GWS-MD106) standard $\chi^2$ spectral fitting was also performed. In this case the spectra were grouped to have a minimum of twenty counts per bin, required for approximately Gaussian statistics. 

The LBG spectra were also searched for the presence of  iron K$\alpha$ line emission by looking for evidence of an excess of counts at rest frame 6.4~keV, as compared to that expected from the best fitting model for each source. None of the LBGs showed significant evidence of Fe K line emission. 

\subsection{C-statistic and $\chi^2$ fitting of individual LBGs}

%%%%%%%%%%%%%%%%%%%%%%%%%
\begin{table*}
\begin{minipage}{176mm}
\begin{center}
\caption{X-ray spectral fits using the C-statistic \citep{cash79} for sources with greater than 10 counts. All errors correspond to the 90 per cent confidence level.
Col.(1): Optical spectral classification; Col.(2): LBG name; Col.(3): Number of counts in spectral fitting; Col.(4): Effective $\Gamma$ for fits assuming only Galactic absorption;  Col.(5): Intrinsic column density, N$_{\rmn H}$, in units of $10^{23}$~cm$^{-2}$, for fits assuming a fixed intrinsic $\Gamma=2$ power-law spectrum with Galactic absorption; Col.(6): Improvement in the C-statistic for an intrinsic $\Gamma=2$ spectrum by allowing the absorption shown in Col.4 compared to zero intrinsic absorption; Col.(7): Intrinsic column density found from simultaneous fitting of whole sample, where the best fitting photon index was $\Gamma=1.96$. For HDF-MD39 and HDF-MD12 (which were excluded from the simultaneous fitting) the results are for a fixed $\Gamma=1.96$ power-law. }
\label{table5}
\begin{tabular}{@{}llcccccccccc@{}}
\hline
Optical & && & Fixed N$_{\rmn{H}}$$=$0& &\multicolumn{2}{c}{Fixed $\Gamma$$=$2} &&&Simultaneous fit, $\Gamma$$=$1.96\\
\cline{4-5} \cline{7-8} \cline{10-11} 
Class & LBG  &  Counts & &$\Gamma$ && N$_{\rmn{H}}$  & $\Delta$C-stat &&&  N$_{\rmn{H}}$ \\
(1)   & (2) & (3) && (4) & & (5)  &(6) &&& (7)   \\
\hline
QSO & HDF-oC34 &198 &&$2.0^{+0.3}_{-0.3}$ && $<0.39$    &  0.1&&&$<0.43$\\
 & HDF-MD39 &3018 &&$1.8^{+0.1}_{-0.1}$ && $0.08^{+0.04}_{-0.04}$    & 11.3 &&&$0.06^{+0.04}_{-0.04}$\\
 & SSA22a-D13 &28 && $1.5^{+0.5}_{-0.5}$&&   $1.52^{+2.06}_{-1.18}$    & 5.5  &&&$1.40^{+2.22}_{-1.11}$\\
 & SSA22a-D12 &24 && $1.1^{+0.6}_{-0.6}$&&  $1.75^{+1.82}_{-1.14}$    & 25.1 &&&$1.67^{+1.91}_{-1.11}$\\
 & GWS-MD106 &111 &&$1.7^{+0.3}_{-0.4}$ && $<0.65$    & 1.2 &&&$<0.68$\\
 & GWS-D54   & 62&& $2.0^{+0.7}_{-0.6}$  &&$<1.06$    & 0.3 &&&$<1.08$\\
 & GWS-oMD13 &21 && $2.5^{+0.8}_{-0.7}$ && $<0.68$    & 0.0 &&&$<0.68$\\
\hline
AGN & HDF-oMD49&21 && $0.2^{+1.50}_{-2.0}$ &&$<13.2$    & 2.4&&&$<13.1$\\
 & HDF-MD12 &481 &&$1.1^{+0.2}_{-0.2}$ && $1.60^{+0.43}_{-0.38}$    & 95.1 &&&$1.56^{+0.40}_{-0.40}$\\
 & Q1422-MD109&15 && $0.6^{+0.7}_{-0.7}$&& $1.77^{+1.41}_{-0.94}$    &  15.8&&&$1.69^{+1.49}_{-0.93}$\\   
 & Q1422-C73  &19 &&$1.0^{+0.7}_{-0.6}$ &&  $2.36^{+1.92}_{-1.41}$    & 1.5 &&&$2.19^{+2.09}_{-1.28}$\\ 
& GWS-M47   &31 && $0.9^{+0.6}_{-0.6}$ &&$1.95^{+2.20}_{-1.37}$    & 7.3 &&&$1.93^{+2.24}_{-1.39}$\\
 \hline
GAL & HDF-D7   &40 && $<-0.5$ &&$>6.70$     & 3.6 &&&$>6.33$\\
 & HDF-C14  &52 &&$1.2^{+1.0}_{-0.8}$  &&$<5.93$    & 0.5 &&&$<5.91$\\
 & HDF-M9   &32 &&$1.3^{+2.5}_{-1.5}$ &&$<5.36$    & 0.4 &&&$<5.35$\\
 & HDF-M35    & 21&&$4.8^{+5.2}_{-7.8}$ && $<340$   & 0.0 &&&$<272$\\
 & GWS-C50   &20 && $1.2^{+1.1}_{-0.9}$&& $<6.45$    & 1.6 &&&$<6.46$\\
\hline
UNCLASSIFIED & HDF-C10  &75 && $2.1^{+0.7}_{-0.6}$ &&$<0.57$    &  0.0&&&$<0.57$\\
 & HDF-MD34 &116 &&$2.0^{+0.4}_{-0.4}$ && $<0.23$    & 0.0 &&&$<0.24$\\
 & GWS-M10   &19 && $2.3^{+2.1}_{-1.7}$ &&$<2.84$    & 0.0 &&&$<2.78$\\
\hline
\end{tabular}
\end{center}
\end{minipage}
\end{table*}
%%%%%%%%%%%%%%%%%%%%%%%%%

For each of the LBGs the data were fitted  in several ways using the C-statistic to assess the level of intrinsic absorption and the effective power-law photon index, $\Gamma$. First, to determine the effective photon index $\Gamma$ of the spectra, the data were fitted with a power-law model with Galactic absorption using the N$_{\rmn{H}}$ values in Table~\ref{table1} and zero intrinsic absorption (column 4 of Table~\ref{table5}). 
Secondly, the data were fitted with a power-law model  absorbed by both an intrinsic column density at the source redshift and a Galactic column density (Table~\ref{table5}, column 5). In this case the photon index was fixed to $\Gamma$=2 in order to constrain the intrinsic N$_{\rmn{H}}$ required to produce the observed spectra. Adopting a fixed $\Gamma$=2 spectrum (the canonical spectrum of unobscured AGN) is an approximation used to assess the intrinsic column density. For sources with a flatter intrinsic spectrum this will result in an overestimation of N$_{\rmn{H}}$. The errors in the fitted values of both $\Gamma$ and N$_{\rmn{H}}$ correspond to the 90 per cent confidence level
for 1 interesting parameter ($\Delta$C$=2.7$). 
As a further measure of the intrinsic absorption, the data were also fitted with fixed $\Gamma$=2 with only Galactic absorption and the improvement in the C-statistic between this fit and the previous fit allowing intrinsic absorption was calculated (Table~\ref{table5}, column 6). Large values of $\Delta$C-stat indicate that the inclusion of an intrinsic column density produced substantial improvement to the fits. 

In addition to the individual C-statistic fits, simultaneous fitting of the sample was also performed to determine the mean photon index of the LBGs as a whole.  In this fit both the photon index and  intrinsic column density were free parameters, but $\Gamma$ was fixed to be the same for all the LBGs. 
HDF-MD39 and HDF-MD12 were excluded from the simultaneous fitting to prevent the signal being dominated by these two very bright LBGs.  The best fitting photon index for the LBG sample was found to be $\Gamma=1.96^{+0.31}_{-0.22}$. The intrinsic column density of each LBG obtained from the spectral fitting is given in 
Table~\ref{table5}, column 7. The values of N$_{\mathrm{H}}$  found from the simultaneous fitting are very similar to those that were found when assuming a fixed $\Gamma=2$ spectrum. 

\begin{figure}
\begin{center}
\includegraphics[width=85mm]{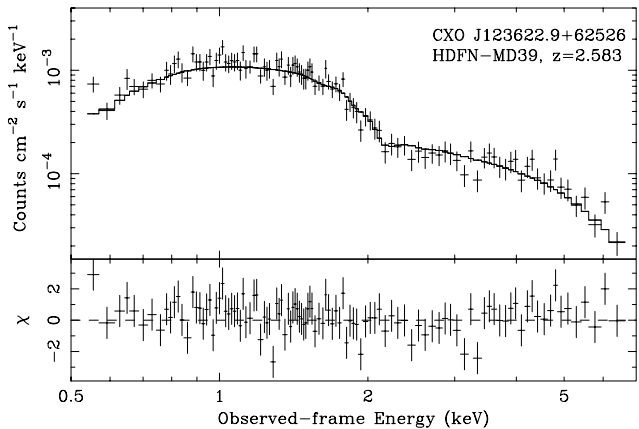}
\caption{The X-ray spectrum of  CXO J123622.9+621526 ($=$HDF-MD39), binned to 20 counts per bin. The fitted model of Galactic absorbed power-law emission with intrinsic $\Gamma=1.80^{+0.09}_{-0.07}$ and negligible intrinsic absorption is shown. Data-to-model residuals are shown in the bottom panel in units of $\sigma$.  
The spectrum is consistent with no intrinsic absorption. }
\label{figure3}
\end{center}
\end{figure}

\begin{figure}
\begin{center}
\includegraphics[width=85mm]{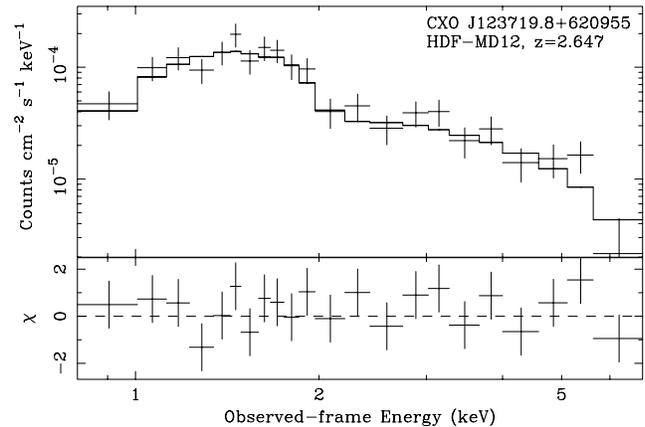}
\caption{The X-ray spectrum of CXO J123719.8+620955 ($=$HDF-MD12),  with a minimum of 20 counts per bin. The fitted model of Galactic absorbed power-law emission with intrinsic $\Gamma=1.92^{+0.51}_{-0.44}$ and intrinsic column density  N$_{\rmn{H}} = 1.65^{+1.15}_{-0.85}\times10^{23}$ is shown (90 per cent confidence range for two interesting parameters).
The bottom panel shows the the data-to-model residuals in units of $\sigma$. 
}
\label{figure4}
\end{center}
\end{figure}

The spectra of the five LBGs with $>$100 counts were fitted using the $\chi^2$ statistic.  Each were fitted with a fixed $\Gamma=2$ power-law spectrum with both intrinsic and Galactic absorption, as was carried out with the C-statistic. The derived N$_{\mathrm{H}}$ values along with the reduced  $\chi^2$ and probability are shown in Table~\ref{table6}, columns 2--4. The three LBGs with $\gtrsim$200 counts  were also fitted with $\Gamma$ as a free parameter. The results are shown in Table~\ref{table6}, columns 5--8. Strong constraints were able to be placed on the values of $\Gamma$ and N$_{\mathrm{H}}$ for HDF-MD39 (Figure~\ref{figure3}) and HDF-MD12 (Figure~\ref{figure4}).

%%%%%%%%%%%%%%%%%%%%%%%%%
\begin{table*}
\begin{minipage}{176mm}
\begin{center}
\caption{X-ray spectral fits using the $\chi^2$-statistic. For all sources, fits were performed assuming Galactic absorption and a fixed intrinsic $\Gamma=2.0$~power-law spectrum. For HDF-oC34, HDF-MD39 and HDF-MD12, additional fits were  performed  where $\Gamma$ was also a free parameter. 
Col.(1): LBG name; Col.(2): Fitted intrinsic column density, N$_{\rmn H}$, in units of $10^{23}$~cm$^{-2}$. Uncertainties refer to $\Delta \chi^2=2.71$, corresponding to the 90 per cent confidence level for one interesting parameter; Col.(3): Reduced $\chi^2$; Col.(4) $\chi^2$ probability; Col.(5): Fitted photon index, $\Gamma$; Col.(6): Fitted intrinsic column density (units of $10^{23}$~cm$^{-2}$). The uncertainties in Cols. 5 and 6 refer to $\Delta \chi^2=4.61$, corresponding to the 90 per cent confidence level for two interesting parameters; Col.(7): Reduced $\chi^2$; Col.(8) $\chi^2$ probability.
}
\label{table6}
\begin{tabular}{@{}lccccccccc@{}}
\hline
& \multicolumn{3}{c}{ABS}  & & \multicolumn{4}{c}{ABS$+$PL} \\ 
\cline{2-4} \cline{6-9}
LBG &  N$_{\rmn{H}}$ & $\chi^{2}_{\nu}$ & Prob. & & $\Gamma$ & N$_{\rmn{H}}$ & $\chi^{2}_{\nu}$ & Prob.\\
(1)   & (2)    & (3) & (4) & & (5) & (6) & (7) & (8)\\
\hline
 HDF-oC34 &  $<0.96$ & 2.10 & 0.05 & & $1.69^{+0.67}_{-0.60}$ & $<1.10$ & 2.25 & 0.047\\
 HDF-MD34 & $<1.55$ & 0.08 & 0.93 & & \\
 HDF-MD12 & $1.78^{+0.56}_{-0.45}$ & 0.76 & 0.76 & & $1.92^{+0.51}_{-0.44}$ & $1.65^{+1.15}_{-0.85}$ & 0.75 & 0.77 \\
 HDF-MD39 & $0.08^{+0.05}_{-0.05}$ & 1.15 & 0.14 & & $1.80^{+0.09}_{-0.07}$ & $<0.04$ & 0.99 & 0.50\\
 GWS-MD106 & $<5.41$ & 0.02 & 0.98 &&  \\
\hline
\end{tabular}
\end{center}
\end{minipage}
\end{table*}
%%%%%%%%%%%%%%%%%%%%%%%%%

The results of the various spectral fits show that the majority  ($\sim 60$~per cent) of the LBGs  have spectra consistent with an unobscured AGN, although several sources (particularly those classed as galaxies optically) are too faint to be able to place reliable constraints. Seven LBGs show evidence of significant obscuration (SSA22a-D13, SSA22a-D12, HDF-MD12, Q1422-MD109, Q1422-C73, GWS-M47 and HDF-D7).  Of the seven LBGs classified as broad line QSOs, two show evidence of intrinsic absorption (SSA22a-D12 and -D13). The brightest LBG in the sample, the QSO HDF-MD39, is well fit by a $\Gamma=1.80$ power-law with no absorption (Table~\ref{table6} and Figure~\ref{figure3}).  Four of the five LBGs classified optically as narrow line AGN exhibit evidence for significant intrinsic absorption, including HDF-MD12 (Figure~\ref{figure4}). The remaining AGN classified LBG, HDF-oMD49, may also be obscured but is too faint to provide reliable constraints. Longer wavelength data lend support to and obscured hypothesis -- the UV spectrum shows evidence of self-absorption, the source has a very bright 24 micron flux ($\sim$300 $\mu$Jy) and a power-law SED in the IR
(C. C. Steidel, priv. communication).

%%%%%%%%%%%%%%%%%%%%%%%%%%%%%%%%%%%%%%
%%%%%%%%%%%%%%%%%%%%%%%%%%%%%%%%%%%%%%%

%%%%%%%%%%%%%%%%%%%%%%%%%%%%%%%%%%%%%%
%%%%%%%%%%%%%%%%%%%%%%%%%%%%%%%%%%%%%%%

\subsection{Simultaneous fitting of LBG subsets}

%%%%%%%%%%%%%%%%%%%%%%%%%
\begin{table}
\begin{center}
\caption{Simultaneous C-statistic fitting for LBG optical classes.  HDF-MD39, HDF-MD12 and sources with $<$10 counts in the spectrum are excluded. Col.(1): Optical spectral classification; Col.(2):  Intrinsic column density for fits with a fixed $\Gamma=1.96$ power-law spectrum; Cols.(3 and 4):
photon index and intrinsic column density  N$_{\rmn H}$ (units of $10^{23}$~cm$^{-2}$)  for two free  parameter fits. All errors correspond to the 90 per cent confidence level.}
\label{table7}
\begin{tabular}{@{}lccccc@{}}
\hline
Optical & \multicolumn{2}{c}{Fixed $\Gamma$$=$1.96} && \multicolumn{2}{c}{Free $\Gamma$ and   N$_{\rmn{H}}$}\\
\cline{2-3} \cline{5-6}
Class   & N$_{\rmn{H}}$ &&& $\Gamma$ & N$_{\rmn{H}}$ \\
(1)   & (2)    &&&(3) & (4) \\
\hline
QSO & $0.26^{+0.22}_{-0.20}$ &&& $1.9^{+0.4}_{-0.3}$ & $0.19^{+0.49}_{-0.19}$\\
AGN & $1.91^{+0.91}_{-0.71}$ &&& $1.8^{+1.1}_{-0.8}$ & $1.71^{+2.07}_{-1.27}$\\
GAL  & $<3.24$ &&& $1.2^{+2.3}_{-0.7}$ & $<5.79$ \\
\hline
\end{tabular}
\end{center}
\end{table}
%%%%%%%%%%%%%%%%%%%%%%%%%

To examine the average spectral properties of the LBGs as a function of optical spectral classification, simultaneous fitting was carried out for separately for objects classified as broad or narrow line AGN, and galaxies.  For each class two simultaneous fits were performed: the first with a fixed $\Gamma=1.96$ power-law spectrum with intrinsic absorption, and the second with a power-law spectrum with intrinsic absorption, allowing $\Gamma$ to be a free parameter. 
Again, the two brightest LBGs were excluded from the fits to prevent the signal being dominated by these individual sources. The results are shown in Table~\ref{table7}.  The mean spectrum of the QSOs is consistent with a $\Gamma\sim1.9$ power-law with at most a small amount of absorption ($1.9^{+4.9}_{-1.9}\times10^{22}$ cm$^{-2}$). The mean spectrum of the AGN shows more absorption (N$_{\mathrm{H}} = 1.7^{+2.1}_{-1.3}\times10^{23}$ for a $\Gamma=1.8$ power-law), consistent with the results of the individual LBGs in \S4.3. The mean spectrum of the galaxy class is poorly constrained.

%%%%%%%%%%%%%%%%%%%%%%%%%%%%%%%%%%%%%%%%%%%%%%%%%%
\section{Stacking of undetected LBGs}

The mean X-ray properties of the LBGs too weak to be directly detected are determined by employing  a stacking technique, shown to be a useful and successful tool with \textit{Chandra}  data (e.g. \citealt{brandt01}; \citealt{horn01};  \citealt{L05}). Previous stacking analysis of $z\sim3$ LBGs in the HDF-N, using the 1~Ms data (N02) and a different sample of galaxies in the 2~Ms data \citep{lehmer05}, have yielded highly significant detections. The emission from  these X-ray weaker LBGs is thought to be dominated by star formation process instead of AGN and the results have, for instance, being used to calculate X-ray derived SFRs. The X-ray derived SFRs from stacking have been shown to be consistent with radio and extinction corrected UV estimates (e.g. \citealt{reddy04}). Here we perform stacking analyses of the undetected LBGs in the HDF-N and  the GWS.

%%%%%%%%%%%%%%%%%%%%%%%%%%%%%%%%%%%%%%%%%%%%%%%%%%
\subsection{Stacking procedure}

The stacking procedure used in this work is identical to that described in \citet{L05} and we include only a brief outline here.  In both the HDF-N and GWS fields all LBGs with an X-ray counterpart are excluded from the stacking, as are those with an unassociated close by X-ray source. Source counts are extracted from the optical LBG positions (corrected to account for the astrometric offset between the X-ray and optical reference frames, \S2.3) using a circular aperture and are summed to find the total counts for the LBGs in the sample. The background counts are estimated using two methods. First, the LBG positions are randomly shuffled by 5--10 arcsec and background counts are extracted from an X-ray source-masked image. Secondly, background counts are extracted from random positions anywhere over the field of view (excluding areas with no exposure or markedly different exposure from that of the LBG positions, in the case of the HDF-N). This is repeated 1000 times for each LBG position. The background counts are then summed and scaled to the same area as the source extraction to find the net source counts.  To identify the X-ray sources for the source-masked images we
 repeated the source detection procedure described above (\S2.3) using a more stringent probability threshold of $10^{-6}$, for which the numbers of spurious sources over the entirety of both survey areas are expected to be small. 
 
\begin{figure*}
\begin{center}
\includegraphics[width=150mm]{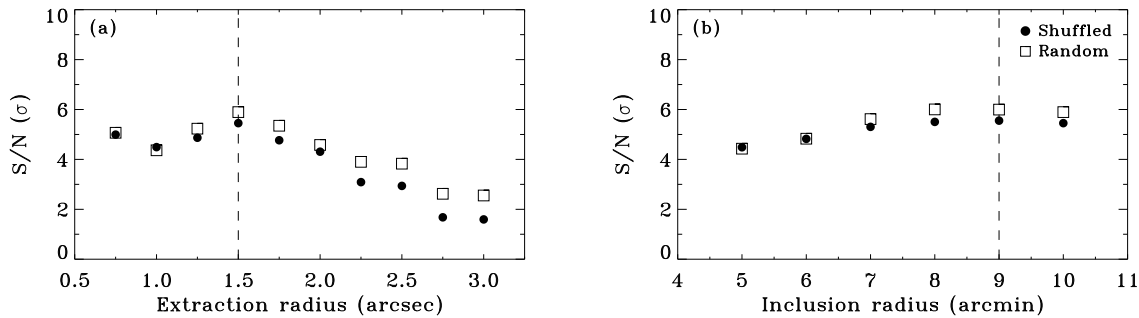}
%\vspace{3.5cm}
\caption{
The signal-to-noise ratio vs. (a) extraction radius in arcseconds and (b) inclusion radius in 
arcminutes for the LBGs in the HDF-N. The results are shown for
two different background methods - shuffled (circles) and random (squares) positions. 
The
signal-to-noise ratio is an inverse measure of the fractional error on
the flux and is given by
$({\rmn{S/{\sqrt{S+B}}}})$, where S and B
are the net source counts and background counts respectively. 
The vertical dashed lines denote our chosen extraction and inclusion 
radii of 1.5 arcsec and 9 arcmin, respectively.
}
\label{figure5}
\end{center}
\end{figure*}

\begin{figure*} 
\begin{center}
\includegraphics[width=150mm]{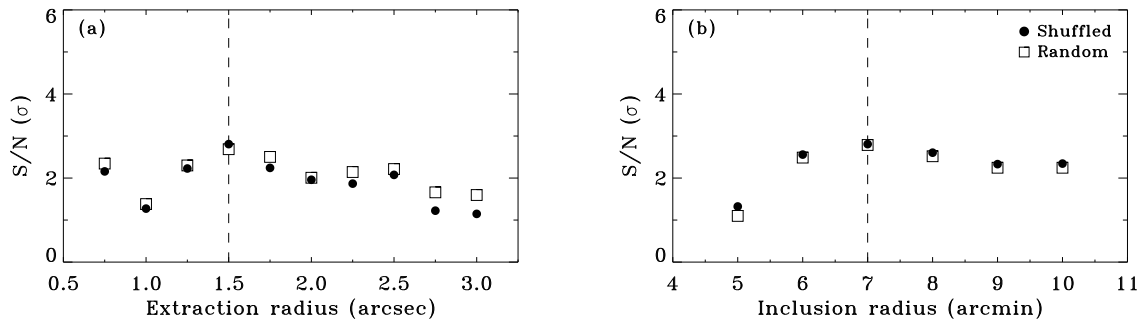}
\caption{The signal-to-noise ratio vs. (a) extraction radius in arcseconds and (b) inclusion radius in 
arcminutes for the LBGs in the GWS. The symbols and axes are the same as for Figure~\ref{figure5}.
The vertical dashed lines denote the chosen extraction and inclusion 
radii of 1.5 arcsec and 7 arc, respectively.}
\label{figure6}
\end{center}
\end{figure*}

The size of the extraction aperture used, as well as the radius from the \textit{Chandra} aim point within which to include LBG positions, affects the strength and accuracy of the stacking signal. We adopt an empirical approach to determining the optimal extraction and inclusion radii by testing 10 fixed extraction radii between 0.75 and 3.0 arcsec and 6 inclusion radii between 5 and 10 arcmin off-axis
and selecting the radii yielding the maximum signal-to-noise ratio.   In the HDF-N we chose an extraction radius of 1.5 arcsec\footnote{We note the N02 contains an error regarding the optimal extraction radius used. The paper states that the optimal extraction radius found was 2.5 arcsec; in fact, 2.5 arcsec was the optimal extraction diameter.} and an inclusion radius of 9 arcmin (Figure~\ref{figure5}). In the GWS an extraction radius of 1.5 arcsec and inclusion radius of 7 arcmin produced the strongest signal-to-noise ratio (Figure~\ref{figure6}). In both fields the shuffled and random background methods produced similar results and in this paper we quote all results using the shuffled positions, which should better account for local variations in the background level.  

In converting the stacking count-rates to fluxes a power-law spectrum with $\Gamma=2.0$ and Galactic absorption was assumed, appropriate for sources with high star formation activity (e.g. \citealt{ptak99}).

%%%%%%%%%%%%%%%%%%%%%%%%%%%%%%%%%%%%%%
%%%%%%%%%%%%%%%%%%%%%%%%%%%%%%%%%%%%%%%

%%%%%%%%%%%%%%%%%%%%%%%%%%%%%%%%%%%%%%
%%%%%%%%%%%%%%%%%%%%%%%%%%%%%%%%%%%%%%%

%%%%%%%%%%%%%%%%%%%%%%%%%%%%%%%%%%%%%%%%%%%%%%%%%%
\subsection{Stacking Results}

\begin{figure}
\begin{center}
\includegraphics[width=65mm]{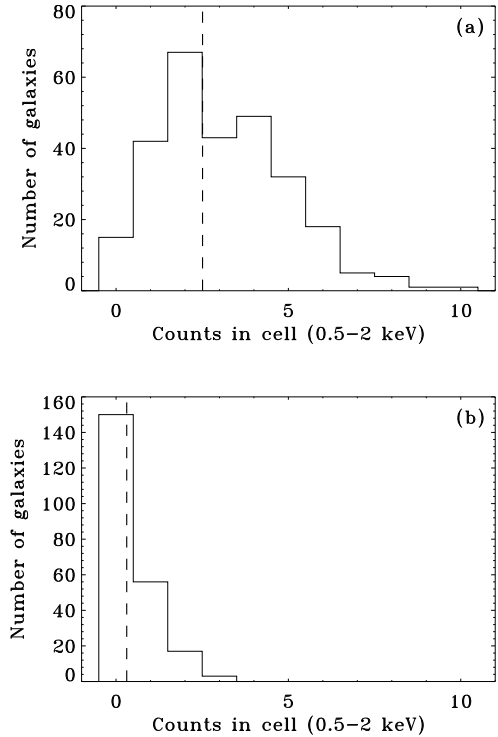}
\caption{Soft band counts distribution for the undetected LBGs in (a) the HDF-N and (b) GWS. The vertical dashed line denotes the mean background counts per extraction cell, derived via the shuffle background method. 
}
\label{figure7}
\end{center}
\end{figure}

The results of the stacking in the HDF-N and GWS are shown in Table~\ref{table8}. Stacking the soft band emission from the 277 undetected LBGs within 9 arcmin from the \textit{Chandra} aimpoint produces  a highly significant detection with a signal-to-noise ratio of 5.9 (where signal-to-noise ratio is defined as ${\rmn{S/{\sqrt{S+B}}}}$ and S and B are the net source counts and background counts, respectively). An average of 0.58 net counts per galaxy are detected, with 2.51 mean background counts in each extraction cell. Figure~\ref{figure7}(a) shows the distribution of total counts (S$+$B) in each LBG extraction cell. The distribution ranges from zero to ten counts per LBG  and is not dominated by a few sources, but rather is well representative of the full sample. The mean flux per HDF-N LBG is $2.36\pm0.43\times10^{-18}$~erg~s$^{-1}$cm$^{-2}$ (0.5--2~keV).  Stacking the soft band emission from 226 LBGs in the GWS also produced a detection, albeit a marginal one. An average of 0.12 net counts per LBG were detected, with 0.31 background counts in each detection cell, resulting in a signal-to-noise ratio of 2.81. The total count in cell distribution for the GWS is shown in Figure~\ref{figure7}(b) and covers a much smaller range than the HDF-N, as would be expected given the shorter exposure time. The mean soft band flux per LBG was found to be $4.34\pm1.54\times10^{-18}$~erg~s$^{-1}$cm$^{-2}$ (0.5--2~keV). The mean flux in the GWS is 84 per cent larger than in the HDF-N, but consistent within the errors at the 1.2$\sigma$ level.  However, a raw comparison of the two results is misleading because of the different flux limits in the GWS and HDF-N. In order to provide a fairer comparison between the two samples the HDF-N stacking was repeated with the inclusion of the six detected LBGs with fluxes below the flux limit in the GWS. The mean flux per LBG in this sample was found to be $3.23\pm0.44\times10^{-18}$~erg~s$^{-1}$cm$^{-2}$ (0.5--2~keV), completely consistent with the GWS result.

Stacking the hard band emission did not result in a significant detection in either field.  

%%%%%%%%%%%%%%%%%%%%%%%%%%%%%%%%%%%%%%
\begin{table*}
\begin{minipage}{176mm}
\caption{Stacking results of undetected LBGs in the HDF-N and GWS, including 
results for sub-samples based on rest-frame
1800~\AA~emission in HDF-N.
Col.(1): Field.
Col.(2): Galaxy sample. 
Col.(3): Observed-frame energy band. 
Col.(4): Number of galaxies included in stacking sample, taking into account rejected galaxies as described in \S2.5. 
Col.(5): Mean $\mathscr{R}$ magnitude. 
Col.(6): Mean redshift. 
Col.(7): Signal-to-noise ratio $[{\rmn{ S/{\sqrt{S+B}}}}]$, where S
and B are the net source and background counts respectively.
Col.(8): X-ray flux per galaxy in units of $10^{-18}~\rmn{erg~cm^{-2}~s^{-1}}$; 0.5--2~keV for soft band, 2--10~keV for hard band.
Col.(9): X-ray luminosity per galaxy in the 2--10~keV band, derived
from soft band flux assuming $\Gamma$=2.0 and Galactic
N$_{\rmn{H}}$. 10--50~keV luminosity is given for hard band, derived
using $\Gamma$=1.4. 
Col.(10): SFR from 2--10~keV luminosity~\citep{ranalli03}. Errors are statistical only.
Col. (11): Ratio of X-ray derived SFR to UV SFR, uncorrected for attenuation.
}
\label{table8}
\begin{tabular}{@{}lllcccrcccc@{}}
\hline
Field &Sample & Band & N & $\langle \mathscr{R} \rangle$ & $\langle z \rangle$ & S/N & F$_{\rmn{X}}$ & L$_{\rmn{X}}$ & $\langle SFR \rangle$ & SFR$_{\rmn{X}}$/SFR$_{\rmn{UV}}^{\rmn{uncor}  }$\\
   &    &      &   &                 &       &     &  & ($10^{41}~\rmn{erg~s^{-1}}$) & (M$_{\sun}~\rmn{yr}^{-1}$) & \\
(1) & (2)  & (3)  & (4)  & (5)  & (6)  & (7)  & (8)  & (9) & (10)&(11)\\
\hline
HDF-N & All undetected & Soft & 277 & 24.82 & 3.00 & 5.9 &$2.36\pm0.43$  & $2.12\pm0.39$ & $42.4\pm7.8$ & $4.1\pm0.8$\\
HDF-N & All undetected  & Hard & 273 & 24.83 & 3.00 & 0.9  &$2.71\pm2.96$ & $2.09\pm2.29$& \ldots& \ldots \\
HDF-N & 22.74 $<\mathscr{R}\le$ 24.65 & Soft & 91  & 24.20 & 2.99 & 3.52 &$ 2.72\pm0.77$ & $2.34\pm0.66$ & $46.8\pm13.2$ & $2.6\pm0.7$\\
HDF-N & 24.65 $<\mathscr{R}\le$ 25.15 & Soft & 96  & 24.93 & 3.01 & 2.98 &$ 2.11\pm0.71$ & $1.90\pm0.64$ & $37.9\pm12.8$ & $4.0\pm1.4$\\
HDF-N & 25.15 $<\mathscr{R}\le$ 25.63 & Soft & 90  & 25.33 & 3.01 & 2.96 &$ 2.31\pm0.78$ & $2.07\pm0.70$ & $41.4\pm15.6$ & $6.3\pm2.4$\\
HDF-N & GWS flux limit$^a$ & Soft & 283 & 24.81 & 3.00 & 7.3 &$3.23\pm0.44$  & $2.78\pm0.38$ & $55.6\pm7.6$ & $5.4\pm0.7$\\
GWS & All undetected & Soft & 226 & 24.70 & 2.94 & 2.81 & $4.34\pm1.54$  & $3.60\pm1.28$ & $72.0\pm25.6$ & $6.4\pm2.3$\\
GWS & All undetected & Hard & 229 & 24.69 & 2.94 & $<$0.0 & \ldots & \ldots & \ldots & \ldots \\
%HDF-N &GWS detect. limit & Hard & 279 & 24.82 & 3.00 & 1.9  &$5.63\pm2.95$ & $4.18\pm2.19$       & \ldots& \ldots \\
\hline
\end{tabular}
$^a$ Includes directly detected LBGs with fluxes below the flux limit of the GWS.
\end{minipage}
\end{table*}
%%%%%%%%%%%%%%%%%%%%%%%%%%%%%%%%%%%%%%

%%%%%%%%%%%%%%%%%%%%%%%%%%%%%%%%%%%%%%
%%%%%%%%%%%%%%%%%%%%%%%%%%%%%%%%%%%%%%%

%%%%%%%%%%%%%%%%%%%%%%%%%%%%%%%%%%%%%%%%%%%%%%%%%%
\subsection{X-ray and UV correlations for stacked LBGs}

The soft band stacking signal in the HDF-N is sufficiently strong to allow the sample to be split into bins according to optical properties and the mean X-ray properties determined for each bin. This has been shown to be very effective in the HDF-N for BBGs at $z\sim1$, allowing correlations between X-ray and rest-frame UV emission to be examined \citep{L05}. In order to achieve reasonable signal-to-noise ratios the LBG sample was split into three bins of approximately 90 galaxies according to $\mathscr{R}$ magnitude, which corresponds to rest-frame $\sim1800$~\AA~emission at $z\sim3$. The stacking was performed according to the procedure described above, using a 1.5 arcsec extraction radius and only including galaxies within 9 arcmin. 

\begin{figure}
\begin{center}
\includegraphics[width=85mm]{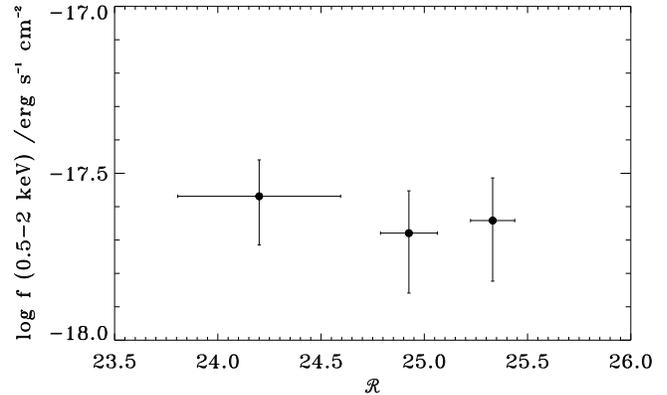}
\caption{X-ray stacking results for undetected LBGs in the HDF-N split into subsets based on $\mathscr{R}$ band magnitude. At $z\sim3$  $\mathscr{R}$ corresponds to rest-frame $\sim1800$~\AA~emission.  The $x$-axis error bars are the standard deviation of values in a given bin and the $y$-axis error bars are the Poisson errors from the stacked soft band counts. There is no correlation between soft X-ray flux and  $\mathscr{R}$ magnitude.}
\label{figure8}
\end{center}
\end{figure}

The subset stacking results are shown in Table~\ref{table8} and Figure~\ref{figure8}. A significant detection was found for the brightest $\mathscr{R}$  bin and marginally significant detections found for the remaining bins. In each case the flux was sufficiently constrained to allow an assessment of any correlations. As can be seen from Figure~\ref{figure8},  within the limited statistics and dynamic range of this LBG sample, we find no evidence for a correlation between  $\mathscr{R}$ magnitude and soft X-ray flux, corresponding to rest-frame 1800~\AA~and 2--8~keV emission.

%%%%%%%%%%%%%%%%%%%%%%%%%%%%%%%%%%%%%%
%%%%%%%%%%%%%%%%%%%%%%%%%%%%%%%%%%%%%%%

%%%%%%%%%%%%%%%%%%%%%%%%%%%%%%%%%%%%%%%%%%%%%%%%%%
\section{Discussion}

We have presented results from Chandra X-ray observations of 6 fields which have been surveyed with deep optical photometric and spectroscopic observations designed to select Lyman break galaxies at $z\sim 3$ \citep{steidel03}. These observations have approximately doubled the number of known X-ray detections of LBGs to a total of 24 (c.f. \citealt{brandt01}; N02; \citealt{nandra05a,lehmer05}). The raw fraction of X-ray detected LBGs in the entire sample is 2~per cent, but it should be borne in mind that many of the Chandra observations presented here are very shallow, and not sufficiently sensitive to probe deep into the X-ray luminosity function. Taking the HDF-N alone, which has the deepest data, we estimate an X-ray detection fraction in LBGs of $3 \pm 1$~per cent, similar to that found by \citet{steidel02}.

It is therefore immediately interesting to note that the fraction of LBGs hosting an AGN is much lower than that of submm selected galaxies \citep{alexander05}, in which more than half the sources are likely to harbour an actively accreting black hole. The importance of this result is at least twofold. Firstly, there remains considerable uncertainty about the contribution of AGN to the bolometric luminosity of FIR selected galaxies at the highest luminosities (e.g. \citealt{veilleux95,farrah02}). No such uncertainty would appear to be pertinent for the LBGs. Secondly, the results for the submm galaxies have been interpreted as a strong linkage between intense star formation and AGN activity at high-redshift (\citealt{alexander05}; see also \citealt{page01,page04}). We find no evidence that LBGs, which are certainly galaxies in which active star formation is occurring, are also preferentially active in nuclear black hole accretion. 

The X-ray--to--optical flux ratios of the LBGs cover almost the full range observed in X-ray selected samples. Objects with the highest X-ray-to-optical ratio (e.g. EXOs; \citealt{koekemoer04}) are not present but this is clearly not surprising given this is an optically selected sample. Most of the X-ray detected LBGs fall within the range expected for AGN. A few objects show lower X-ray-to-optical flux ratios, but four of these are clearly AGN also, based either on their optical spectroscopic properties, X-ray luminosity, or X-ray spectral properties. LBGs hosting AGN therefore display a range in X-ray to-optical flux ratio of at least two orders of magnitude. 

There are two X-ray detected LBGs, HDF-M9 and HDF-35 which exhibit very low X-ray-to-optical flux ratios, typical of starburst galaxies. Furthermore, both objects have soft X-ray spectra, low X-ray luminosity ($\sim 1-3 \times 10^{42}$~erg s$^{-1}$) and  have no indication of AGN activity in their optical spectra. It is possible that their X-ray emission is simply due to low luminosity AGN activity. It also seems quite plausible, however, that the X-rays from these two objects are powered by intense star forming activity, with the emission being predominantly from X-ray binaries and hot gas \citep{david92}. Using the standard conversions between X-ray luminosity and star-formation rate (N02; \citealt{ranalli03,grimm03}), we infer star formation rates for the two objects of 292$^{+111}_{-83}$~M$_{\rm \odot}$ yr$^{-1}$ (HDF-M9) and 521$^{+189}_{-142}$~M$_{\rm \odot}$ yr$^{-1}$ (HDF-M35). Clearly these objects would then be considered quite extreme, with SFRs comparable to bright submm selected galaxies (e.g. \citealt{ivison00}) or Hyper luminous IRAS galaxies \citep{rowan00}. These high SFRs should be compared to the typical, extinction-corrected average values for LBGs, which we find in this work to be $\sim 40-50$~M$_{\rm \odot}$ yr$^{-1}$ (see also \citealt{steidel99}; N02). Given the large sample under consideration, and the fact that we expect a wide range of SFRs in the LBGs, it seems quite reasonable that there are extreme objects with SFRs a factor of several higher than the mean. Subsequent analysis of {\it Spitzer} IRAC and MIPS 24$\mu$m  observations yielded SFRs of $\sim$170~M$_{\rm \odot}$ yr$^{-1}$ (HDF-M9) and $\sim$360~M$_{\rm \odot}$ yr$^{-1}$ (HDF-M35) and supports a star formation hypothesis for both these LBGs  (C. C. Steidel, priv. communication).

Whatever the origin of the X-ray emission in these two objects, our analysis has revealed new information about the AGN population at high-redshift.  We find that the X-ray detected LBGs display a wide range of luminosities from $\log L_{\rm X} = 42-45$. This is typical of the AGN population at lower redshift (e.g. \citealt{cowie02,rosati02}). The optical spectral classifications of the X-ray detected objects are quite diverse, comprising approximately 1/3 broad line AGN, 1/3 narrow line AGN and 1/3 with spectra typical of normal star forming LBGs. These proportions are again very similar to AGN in X-ray selected samples in general \citep{barger03}. The implication is that the range of properties of AGN in the LBG population at $z\sim 3$ are very similar to those at lower redshift. In particular, our results support the idea that there is a large population of low luminosity ``Seyfert'' level AGN of all optical types at $z=3$  \citep{steidel02,nandra05a}.

%%%%%%%%%%%%%%%%%%%%%%%%%%
\begin{table}
\begin{center}
\caption{X-ray luminosities of the LBGs. Col.(1): LBG name; Col.(2): Optical classification; Col.(3): Rest-frame 2--10~keV luminosity, in units of $10^{43}$~erg~s$^{-1}$; Col(4): Absorption-corrected rest-frame 2--10~keV luminosity, for sources with evidence for absorption in their spectrum.}
\label{table9}
\begin{tabular}{@{}lccc@{}}
\hline
LBG & Optical & L$_{2-10~\rmn{keV}}$ & L$_{2-10~\rmn{keV}}$  \\
 & type & uncorrected & corrected \\
(1)   & (2) & (3) & (4)            \\
\hline 
HDF-D7   & GAL & $0.42^{+0.09}_{-0.08}$&$11.45^{+6.89}_{-11.04}$ \\
HDF-C14  & GAL &$0.64^{+0.15}_{-0.12}$&$1.02^{+1.48}_{-0.32}$ \\
HDF-oC34 & QSO &$5.05^{+0.46}_{-0.42}$&\\
HDF-C10  &\ldots &$1.13^{+0.19}_{-0.16}$&\\
HDF-MD34 &\ldots &$2.24^{+0.27}_{-0.25}$&\\
HDF-oMD49 & AGN& $0.67^{+0.22}_{-0.17}$&$4.43^{+86.63}_{-3.22}$\\
HDF-MD12 & AGN &$6.84^{+0.44}_{-0.42}$&$16.81^{+11.24}_{-6.80}$ \\
HDF-MD39 & QSO &$49.79^{+1.05}_{-1.00}$& \\
HDF-M35  & GAL &$0.17^{+0.07}_{-0.05}$&\\
HDF-M9   & GAL &$0.26^{+0.10}_{-0.07}$ & $0.44^{+5.92}_{-1.79}$ \\
\hline
Q1422+2309b &QSO & $10.75^{+5.82}_{-3.96}$ &\\
Q1422-MD109 &AGN &$8.31^{+3.56}_{-2.58}$ &$19.43^{+9.08}_{-6.79}$ \\
Q1422-C73  & AGN &$17.83^{+6.45}_{-4.87}$ &$47.64^{+24.20}_{-6.79}$ \\
\hline
SSA22a-D13 & QSO &$14.13^{+3.93}_{-3.14}$ &$28.90^{+14.03}_{-9.02}$\\
SSA22a-M8  &\ldots&$1.75^{+1.19}_{-0.75}$ &\\
SSA22a-D12 & QSO &$7.51^{+2.39}_{-1.85}$ & $16.58^{+8.19}_{-5.13}$\\
SSA22a-MD14 &GAL & $1.93^{1.31}_{-0.83}$ &\\
SSA22a-M14 & GAL &$2.76^{+1.49}_{-1.02}$ &\\
\hline
GWS-MD106 & QSO &$16.78^{+2.20}_{-1.96}$ &$21.05^{+7.77}_{-2.71}$\\
GWS-D54   & QSO &$12.09^{+2.37}_{-2.01}$ &\\
GWS-M47   & AGN &$4.88^{+1.50}_{-1.17}$ &$12.50^{+5.63}_{-6.34}$ \\
GWS-M10   &\ldots&$2.31^{+0.88}_{-0.66}$ & \\
GWS-oMD13 & QSO &$4.26^{+1.26}_{-0.99}$ &\\
GWS-C50   & GAL &$2.36^{+0.95}_{-0.70}$ &$5.99^{+5.91}_{-3.95}$ \\
\hline
\end{tabular}
\end{center}
\end{table}
%%%%%%%%%%%%%%%%%%%%%%%%%%

X-ray spectral analysis of the LBGs shows evidence for significant absorption in several objects. This may be considered surprising given both the limited photon statistics in the spectra, and the fact that at $z\sim 3$ we sample the spectra only at $E>2$~keV in the rest frame, so that only the heaviest absorption is detectable.   Formally, we find evidence for absorption in about half the sample, but given the above considerations the true fraction could be higher. Splitting the objects according to optical spectroscopic class and fitting simultaneously, we find the clearest evidence for absorption in the narrow line objects, consistent with standard unification schemes where the same material obscures both the X-rays and the optical BLR \citep{antonucci85}. The columns are typically $10^{23}$~cm$^{-2}$,  a little lower than those for type II Seyferts in the local universe \citep{awaki91}. Once the X-ray fluxes are corrected for absorption (Table~\ref{table9}), a number of the narrow line objects fall into the luminosity and obscuration range where they would be classified as candidate ``type II QSOs''. For example, four objects have $L_{\rm X}>10^{44}$ and $N_{\rm H} > 10^{22}$~cm$^{-2}$. These have been used as typical dividing lines (e.g. \citealt{mainieri02,brusa05}) although it should be noted that absorbing columns of $10^{22}$~cm$^{-2}$  should be considered on the very low side for bona fide Seyfert 2s \citep{awaki91}. The nature and origin of the absorption in these lightly-obscured objects, and hence in the LBGs, remains uncertain and could easily be related to galactic-scale dust and gas rather than a standard torus \citep{maiolino95}. It should further be noted that at least two objects show QSO luminosity and heavy obscuration in excess of $10^{23}$~cm$^{-2}$, but have broad emission lines in their optical spectra. These would be classified as candidate type II QSOs based solely on their X-ray properties (e.g. \citealt{mainieri02}) but are clearly not type IIs by definition. How these objects fit into the unification schemes is currently unclear, but it is possible either that the obscuration is purely nuclear, affecting the X-rays only, or that the torus is relatively warm and ionized, and hence dust-free. 

Taking a naive flux-luminosity  conversion we find that the narrow line AGN sample is less luminous in the X-rays than the broad line QSOs  (Table~\ref{table10}). This effect seems to be primarily due to the more common incidence of heavy absorption in the AGN. After correction for obscuration we find the average luminosity of the broad and narrow line objects to be very similar. Objects classified as galaxies from optical spectroscopy (i.e. without AGN signatures) are, on the other hand, roughly an order of magnitude less luminous in the X--ray than optically-identified AGN (Table~\ref{table10}). This conclusion holds even after absorption is accounted for. The galaxy-classified objects tend not to show evidence for absorption (with the clear exception of HDF-D7), but small photon statistics prevent definitive conclusions. It is therefore possible that large-scale dust obscured both the broad and narrow lines in these objects. It seems more likely based on our results, however, that the lack of optical AGN signatures is due to their low intrinsic luminosity, so that the galaxy light dominates over the AGN in the optical \citep{moran02}
 
%%%%%%%%%%%%%%%%%%%%%%%%%%
\begin{table}
\begin{center}
\caption{Mean 2--10~keV luminosity of LBG optical classes. Col.(1): Optical spectral classification.
Col.(2): Mean 2--10~keV luminosity of spectral group, uncorrected for X-ray absorption.
Col.(3): Mean 2--10~keV luminosity of spectral group, corrected for X-ray absorption.
}
\label{table10}
\begin{tabular}{@{}lcc@{}}
\\
\hline
Optical & Uncorrected L$_{2-10 keV}$ & Corrected L$_{2-10 keV}$ \\
Class   &   ($10^{43}~\textrm{erg~s}^{-1}$) &  ($10^{43}~\textrm{erg~s}^{-1}$)\\
(1)   & (2) & (3)\\
\hline
QSO & $15.05\pm14.69$ & $18.56\pm15.03$ \\
AGN & $7.71\pm6.35$ &  $20.16\pm16.38$ \\
GAL & $1.22\pm1.09$ & $3.39\pm4.06$\\
\hline
\end{tabular}
\end{center}
\end{table}
%%%%%%%%%%%%%%%%%%%%%%%%%%

A final note regarding the X-ray spectral properties of the directly detected LBGs is that their mean continuum spectral index of $\Gamma=1.96$ is remarkably similar to that of local Seyferts \citep{nandra94}. We therefore find no evidence for evolution of the continuum-generating mechanism with redshift. 

The mean properties of the LBGs not directly detected in the deep HDF-N and GWS observations have been determined by stacking. This technique has been successfully applied in the past to constrain LBG X-ray emission which is thought to be dominated by star-forming processes, rather than AGN activity (\citealt{brandt01}; N02; \citealt{lehmer05}). The other data sets in this study are too shallow to constrain the properties of such sources and, indeed even the 200~ks GWS observation is of limited value. Analysis in the deeper HDF-N field shows a population of LBGs, the majority of which are AGN, with X-ray fluxes just below the flux limit of the GWS observation. When these sources are included in the HDF-N stacking sample they dominate the average stacked flux. This shows that there is a clear risk of contamination in the GWS signal by X-ray sources that are just below the flux limit for direct detection, and whose emission is likely dominated by black hole accretion rather than star formation. The best estimate of the stacked flux in the GWS, while poorly constrained, is about $80$~per cent higher than that of the HDF-N, probably due to this contamination. This shows the limited usefulness of the 200~ks GWS observations in determining the properties of star-forming LBGs and henceforth we restrict our discussion to the HDF-N.

We find a highly significant signal in the soft band when stacking all 277 undetected LBGs in the HDF-N. There is no detection in the hard band. This is primarily due to the lower ACIS sensitivity in the hard band, but it shows at least that the stacked objects are consistent with a relatively soft X-ray spectrum expected from star formation. The inferred flux per object is entirely consistent with previous stacking analysis, and specifically that inferred by N02 using the 1~Ms data. This is important because it shows that the signal detected by N02 was not dominated by a few bright, AGN-dominated sources. It should be noted, however, that at least one clear X-ray AGN (HDF-C14) has been detected in the 2~Ms
data that was not detected in 1~Ms. The case of HDF-oMD49 is particularly interesting because this object has already been identified by \citet{steidel02} as an AGN based on optical spectroscopic signatures, but was not detected in the 1~Ms Chandra data. While we have now found a detection in the 2~Ms observation this highlights the very wide range of X-ray--to--optical flux ratio in the LBG AGN discussed above. 

The mean star formation rate inferred from the X-rays for the LBGs is found to be $42\pm 8$~\Msolar 
yr$^{-1}$. Using the ratio of the X-ray derived SFR (which should be roughly independent of extinction) to the UV derived SFR (uncorrected for extinction),  the inferred UV extinction correction at 1800~\AA~is found to be $4.1\pm 0.8$ (Table~\ref{table8}). This is in good agreement with the extinction correction inferred previously by N02, and with that determined for galaxies at $z\sim 2$ \citep{reddy04,reddy05}. It also agrees with, and validates, the extinction corrections employed by \citet{steidel99}, which were calculated using the \citet{calzetti97}  reddening law with $E(B-V)=0.15$, typical for the LBGs.  It also agrees well with the analysis of \citet{pettini98}, which used H$\beta$~luminosity as a comparison SFR measure, and the UV/X-ray analysis of \citet{seibert02}. 
This extinction correction is however on the lower end of an estimate by \citet{vijh03}, who found the UV attenuation factor for a large sample of $z\sim3$ LBGs to be luminosity dependent and in the range 5.9 to 18.5.

Our sample is now large enough, and the stacking signal strong enough to move beyond the average properties. The LBG sample has therefore been split according to optical magnitude in a manner similar to that employed by \citet{L05} for $z\sim1$ galaxies. With the limited statistics of only three bins and large error bars we find no evidence for a direct correlation between $\mathscr{R}$ and soft X-ray flux (Figure~\ref{figure8} and Table~\ref{table8}), which corresponds to rest-frame 1800~\AA~and 
2--10~keV emission at $z\sim3$. This is possibly in contrast to that seen for $z\sim1$ star forming galaxies where a linear relation between L$_{\mathrm X}$ and L$_{\mathrm {UV}}$ was found \citep{L05}. While it is difficult to draw conclusions from this data, one interpretation of this result is that at the higher SFRs under consideration here, the effects of dust attenuation are stronger and act to remove the direct correlation between UV luminosity and SFR \citep{adelberger00}.   As can be seen in Table~\ref{table8} the extinction estimates actually decrease for the more UV luminous LBGs, although within the large errors they are consistent with being constant. This suggests that UV attenuation may not be a direct function of SFR over a small range in SFRs.

\section{Summary}

Using six  \citet{steidel03} LBG survey fields with \chandra/ACIS imaging, the X-ray properties of a large sample of UV-selected $z\sim3$ LBGs have been examined. The X-ray data were used to identify and study AGN within the sample, as well as to provide a high-energy perspective on the star formation in the galaxies that do not harbour an AGN.  The main results are:
\begin{enumerate}

\item 24 LBGs have been detected, approximately doubling the number of known X-ray detections. The increased sample of X-ray detected LBGs results from new analysis of \chandra~archive data covering four of the \citet{steidel03}  survey fields and the doubling of the exposure of the \chandra~data in the HDF-N, compared to the analysis of N02. The AGN fraction in LBG surveys is approximately 3 per cent, much lower than submm galaxies.

\item Around $1/3$ of the X-ray detected LBGs were identified as broad line AGN, $1/3$ as narrow line AGN and $1/3$ as normal star forming galaxies at $z\sim3$. The X-ray luminosities of the LBGs ranged from $1.5\times10^{42}$ to $5\times10^{44}$~erg~s$^{-1}$ (2--10~keV),  therefore spanning  Seyfert to quasar luminosities. The range in luminosity and the breakdown of spectral types is similar to that seen in X-ray surveys at lower redshift. Deep and comprehensive optical spectroscopy is therefore a reasonably good way of identifying AGN at $z$$\simeq$3 \citep{steidel02}, in this case missing about 30 per cent of the population. 

\item The X-ray to optical flux ratio of the LBGs covers two orders of magnitude, ranging from classic AGN to low luminosity AGN and starburst values. Two LBGs (HDF-M9 and HDF-M35) have low X-ray to optical flux ratio, soft X-ray emission, low X-ray luminosity and may be powered in the X-ray not by an AGN but by starburst activity. If powered solely by star formation, these LBGs have X-ray derived SFRs in the range 300 to 500 \Msolar~yr$^{-1}$ and would represent the upper tail of the LBG SFR distribution. Strong MIPS 24$\mu$m detections and similar FIR derived SFRs lend support to the star formation hypothesis.  

\item The X-ray spectra were analyzed for each of the LBGs with greater than 10 counts in the extracted spectrum. The mean photon index of the X-ray detected LBGs,  allowing for intrinsic absorption in each of the sources, was found to be $\Gamma=1.96^{+0.31}_{-0.22}$. 

\item Significant obscuration was detected in 40 per cent of the LBGs. All but one of the narrow line AGN have N$_{\mathrm H} > 10^{23}$~cm$^{-2}$. Four luminous, obscured type 2 QSO candidates are found after the X-ray luminosities  have been corrected for the observed column densities, although definitive confirmation of the type II nature of these objects is impossible without rest-frame optical Balmer line spectroscopy. 

\item A naive conversion from flux gives a mean luminosity for the narrow line AGN LBGs approximately half that of the QSOs. However, as a group the narrow line AGN  were found to have significantly higher N$_{\mathrm H}$ than the QSO group and after correcting the luminosities for absorption, the mean luminosity of the AGN and QSO classes were similar. The LBGs classified optically as normal galaxies are approximately an order of magnitude less luminous than the optically identified AGN. The optical light in these weak AGN is probably dominated by the host galaxy. 

\item Stacking the soft band flux of 277 undetected LBGs in the 2~Ms HDF-N produces a highly significant detection. The resulting average luminosity of L$_{2-10~\mathrm{keV}} =  2.12\pm0.39 \times 10^{41}$~erg~s$^{-1}$  and X-ray derived SFR of $42\pm8$~\Msolar~yr$^{-1}$ are in excellent agreement with that  found by N02 using the 1~Ms data. The implied UV extinction correction is a factor $4.1 \pm 0.8$, consistent with that found in several previous studies. Stacking the hard band flux did not produce a detection.

\item Stacking the soft band flux of 226 undetected LBGs in the 200~ks GWS produced a marginally significant,  3.3$\sigma$ detection. The derived average flux, luminosity and SFR were larger than, but consistent with those found for the HDF-N. There is some evidence that the elevated flux is due to the presence of AGN within the  stacking sample. 

\item Splitting the LBG sample into three subsets based on observed frame $\mathscr{R}$ magnitude and stacking the X-ray flux produced marginally significant results in each bin, allowing the correlation between X-ray and UV emission of LBGs to be examined. Unlike for UV-selected galaxies at $z\sim1$, there is no evidence for a correlation between rest-frame X-ray and UV flux.

\end{enumerate}

%%%%%%%%%%%%%%%%%%%%%%%%%%%%%%%%%%%%%%%%%%%%%%%%%%

%%%%%%%%%%%%%%%%%%%%%%%%%%%%%%%%%%%%%%%%%%%%%%%%%%
\section*{Acknowledgements}
We acknowledge the financial support of PPARC (ESL) and the Leverhulme Trust (KN). This work has made us of data from the \textit{Chandra} archive hosted at CfA. We are grateful to those who built and operate the \textit{Chandra} X-ray Observatory.

%%%%%%%%%%%%%%%%%%%%%%%%%
\begin{landscape}
\begin{table}
\caption{X-ray detected LBGs in the \citet{steidel03} fields. 
Col.(1): \textit{Chandra} designation (J2000).
Col.(2): LBG survey name from \citet{steidel03}, except HDF-MD39 which is from new LBG survey.
Col.(3): $\mathscr{R}$ magnitude.
Col.(4): Spectroscopic redshift.
Col.(5): Positional offset in arcseconds after applying astrometric shift as described in \S2.3.
Col.(6): Full band background subtracted photons detected in 90 per cent EEF area. 
Col.(7): Lowest false detection probability found for the four detection bands. Probability of 10$^{-8}$ is assigned if Poisson probability is less than this value.
Col.(8): Full band flux (all fluxes $10^{-16}~\rm{erg~cm^{-2}~s^{-1}}$). 
Col.(9): Soft band flux. 
Col.(10): Hard band flux. 
Col.(11): Rest-frame 2--10~keV luminosity in units of $10^{42}~\rm{erg~s^{-1}}$, converted from soft band flux.
Col.(12): Hardness ratio, ${\rmn{ HR = (H-S)/(H+S)}}$, where H and S are the net hard and soft band counts, corrected to on-axis values.
Col.(13): Optical spectral classification.
Col.(14): Previous identification as X-ray detected $z$$\sim$3 LBGs: (a) N02, (b) \citealt{basu-zych04},  (c) \citealt{nandra05b}.
}
\label{table3}
\begin{tabular}{@{}llcccccccccccc@{}}
\hline
X-ray ID & LBG  & $\mathscr{R}$ & $z$ & offset & Net Counts & $p_{\rm{min}}$ & F$_{0.5-10\rm{keV}}$ & F$_{0.5-2\rm{keV}}$ & F$_{2-10\rm{keV}}$ & L$_{2-10\rm{keV}}$ & HR & Optical & Ref \\
CXO     & Name &               &     & (\arcsec) & (0.5-7~keV)  &                &      &                     &                    &  & & Type &  \\
(1) & (2)  & (3)  & (4)  & (5)  & (6)  & (7)  & (8)  & (9)  & (10)  & (11) & (12) & (13) & (14) \\
\hline
J123618.4+621139  &  HDF-D7 &  24.55 & 2.394 & 0.66 &    9.37$^{+ 6.81}_{- 5.71}$ & $9.3\times10^{-5}$ & $<2.00$     & $<0.40$ &    2.25$^{+ 0.49}_{- 0.41}$ &    4.18$^{+ 0.91}_{- 0.76}$ & 1.00 & GAL \\ 
J123622.5+621306 &         HDF-C14 & 24.92 & 2.981 & 0.34 &   26.37$^{+ 7.38}_{- 6.30}$ & $1.0\times10^{-8}$ &    3.66$^{+ 0.68}_{- 0.58}$ &    0.72$^{+ 0.17}_{- 0.14}$ &    2.45$^{+ 0.64}_{- 0.52}$ &    6.68$^{+ 1.55}_{- 1.28}$ & -0.25 & GAL  \\ 
J123633.4+621418 &   HDF-oC34  & 25.32 & 3.413 & 0.29 &  184.95$^{+14.96}_{-13.92}$ & $1.0\times10^{-8}$ &   21.66$^{+ 1.67}_{- 1.55}$ &    4.17$^{+ 0.38}_{- 0.35}$ &    9.81$^{+ 1.42}_{- 1.25}$ &   54.49$^{+ 4.95}_{- 4.55}$ & -0.42 & QSO   & a \\ 
J123655.8+621201 &   HDF-C10 & 24.36 & \ldots & 0.27 &   58.14$^{+ 9.35}_{- 8.28}$ & $1.0\times10^{-8}$ &    6.84$^{+ 0.93}_{- 0.82}$ &    1.28$^{+ 0.22}_{- 0.19}$ &    3.02$^{+ 0.73}_{- 0.60}$ &   12.44$^{+ 2.09}_{- 1.81}$ & -0.42 & \ldots  & a \\ 
J123702.6+621244 &   HDF-MD34  & 25.32 & \ldots & 0.33 &  104.73$^{+11.81}_{-10.75}$ & $1.0\times10^{-8}$ &   12.19$^{+ 1.24}_{- 1.13}$ &    2.34$^{+ 0.29}_{- 0.26}$ &    4.70$^{+ 0.94}_{- 0.79}$ &   22.73$^{+ 2.78}_{- 2.49}$ & -0.48 & \ldots  & a \\ 
J123704.2+621446 &   HDF-oMD49 & 24.78 & 2.211 & 0.30 &    8.50$^{+ 4.97}_{- 3.83}$ & $1.2\times10^{-4}$ &    1.27$^{+ 0.42}_{- 0.32}$ & $<0.30$ & $<2.02$  &    1.51$^{+ 0.50}_{- 0.38}$ & \ldots & AGN \\ 
J123719.8+620955 &   HDF-MD12  & 24.84 & 2.647 & 0.13 &  407.31$^{+22.88}_{-21.85}$ & $1.0\times10^{-8}$ &   52.48$^{+ 2.51}_{- 2.40}$ &    7.99$^{+ 0.52}_{- 0.49}$ &   34.63$^{+ 2.48}_{- 2.32}$ &   55.02$^{+ 3.55}_{- 3.34}$ & -0.14 & AGN  & a \\        
J123622.9+621526 &         HDF-MD39  & 20.48 & 2.583 & 0.16 & 3170.66$^{+57.50}_{-56.49}$ & $1.0\times10^{-8}$ &  381.09$^{+ 6.87}_{- 6.74}$ &   74.11$^{+ 1.54}_{- 1.51}$ &  159.76$^{+ 5.62}_{- 5.43}$ &  501.62$^{+10.43}_{-10.22}$ & -0.45 & QSO  \\ 
J123645.0+621653 &         HDF-M35 & 24.05 & 3.229 &  1.15 &    2.75$^{+ 5.09}_{- 3.95}$ & $6.7\times10^{-5}$ &   $<1.97$ &    0.25$^{+ 0.09}_{- 0.07}$ & $<2.68$ &    2.89$^{+ 1.05}_{- 0.79}$ & -1.00 & GAL  \\ 
J123651.5+621041 &         HDF-M9 &  24.41 & 2.975 & 0.49 &    5.43$^{+ 5.56}_{- 4.43}$ & $4.8\times10^{-5}$ &    $<1.74$  &    0.17$^{+ 0.07}_{- 0.05}$ & $<2.35$ &    1.62$^{+ 0.62}_{- 0.46}$ & -1.00 & GAL  \\ 
\hline
J142440.7+225542 & Q1422+2309b & 22.09 & 3.630 & 0.09 & 7.18$^{+ 3.96}_{- 2.76}$ & $1.0\times10^{-8}$ &   34.09$^{+16.87}_{-11.78}$ &    6.90$^{+ 3.73}_{- 2.54}$ & $<30.72$ &  105.89$^{+57.24}_{-38.98}$ & -1.00 & QSO \\
J142442.7+225446 & Q1422-MD109    & 23.69 & 2.229 & 0.21 & 14.47$^{+ 4.97}_{- 3.83}$ & $1.0\times10^{-8}$ &   68.14$^{+22.57}_{-17.38}$ &   10.83$^{+ 4.63}_{- 3.36}$ & $<30.70$ &   51.28$^{+21.94}_{-15.92}$ & -1.00 & AGN  \\
J142446.5+225545 & Q1422-C73  & 24.88 & 3.376 & 0.18 & 18.41$^{+ 5.44}_{- 4.32}$ & $1.0\times10^{-8}$ &   86.30$^{+24.73}_{-19.61}$ &   13.95$^{+ 5.05}_{- 3.82}$ &   49.45$^{+29.65}_{-19.57}$ &  178.97$^{+64.81}_{-48.93}$ & -0.36 & AGN\\
\hline
J221722.3+001640 & SSA22a-D13  & 20.84 & 3.353  & 0.68 & 25.64$^{+ 6.17}_{- 5.06}$ & $1.0\times10^{-8}$ &   58.43$^{+13.87}_{-11.38}$ &   12.99$^{+ 3.61}_{- 2.88}$ & $<13.72$ &  165.26$^{+45.90}_{-36.61}$ & -1.00 & QSO \\
J221725.2+001156 & SSA22a-M8   & 24.72 & \ldots & 0.99 & 3.73$^{+ 3.40}_{- 2.15}$ & $2.0\times10^{-6}$ &  $<10.34$ &    1.97$^{+ 1.34}_{- 0.85}$ & $<13.80$ &   19.14$^{+13.00}_{- 8.24}$ & -1.00 & \ldots  \\
J221736.6+001622 & SSA22a-D12  & 21.61 & 3.084  & 0.45 & 22.37$^{+ 5.87}_{- 4.76}$ & $1.0\times10^{-8}$ &   39.24$^{+10.02}_{- 8.12}$ &    6.87$^{+ 2.19}_{- 1.70}$ &   24.97$^{+12.35}_{- 8.62}$ &   71.49$^{+22.75}_{-17.67}$ & -0.32 & QSO\\
J221738.1+001344 & SSA22a-MD14 & 24.14 & 3.094  & 0.81 & 4.16$^{+ 3.40}_{- 2.15}$ & $5.8\times10^{-6}$ & $<10.29$ &    2.26$^{+ 1.54}_{- 0.97}$ & $<16.39$ &   23.70$^{+16.11}_{-10.21}$ & -1.00 &  GAL  \\
J221739.1+001331 & SSA22a-M14  & 25.47 & 3.091  & 0.58 & 5.85$^{+ 3.78}_{- 2.58}$ & $4.8\times10^{-7}$ &   10.26$^{+ 5.55}_{- 3.78}$ & $<1.89$ &   14.48$^{+ 9.84}_{- 6.24}$ &   27.59$^{+14.92}_{-10.16}$ & 1.00 & GAL  & b\\ 
\hline
J141747.4+523510 & GWS-MD106 &  22.64 & 2.754  & 0.82 &   87.04$^{+10.89}_{- 9.83}$ & $1.0\times10^{-8}$ &   58.36$^{+ 6.55}_{- 5.91}$ &   17.99$^{+ 2.36}_{- 2.10}$ &   39.88$^{+ 8.23}_{- 6.90}$ &  140.93$^{+18.51}_{-16.45}$ & -0.42 &  QSO  & c \\
J141755.5+523532 & GWS-D54   &  22.77 & 3.199  & 1.18 &   40.30$^{+ 8.33}_{- 7.25}$ & $1.0\times10^{-8}$ &   27.35$^{+ 4.30}_{- 3.74}$ &    7.87$^{+ 1.54}_{- 1.30}$ &   24.08$^{+ 5.99}_{- 4.88}$ & 89.12$^{+17.48}_{-14.78}$ & -0.28 &   QSO  & c \\
J141757.4+523106 & GWS-M47   &  24.30 & 3.026  & 0.75 &   29.41$^{+ 6.63}_{- 5.53}$ & $1.0\times10^{-8}$ &   18.77$^{+ 4.02}_{- 3.35}$ &    4.07$^{+ 1.25}_{- 0.98}$ &   18.08$^{+ 6.25}_{- 4.77}$ &   40.31$^{+12.36}_{- 9.67}$ & -0.10 &  AGN  & c \\
J141800.9+522325 & GWS-M10   &  25.31 & \ldots & 0.96&    7.69$^{+ 4.71}_{- 3.55}$ & $4.7\times10^{-6}$ & $<5.73$ &    2.47$^{+ 0.94}_{- 0.70}$ & $<11.47$ &   23.94$^{+ 9.12}_{- 6.80}$ & -1.00 &  \ldots  & c \\
J141801.1+522941 & GWS-oMD13 &  23.33 & 2.914  & 0.80 &   18.88$^{+ 5.56}_{- 4.43}$ & $1.0\times10^{-8}$ &   12.05$^{+ 3.35}_{- 2.67}$ &    4.37$^{+ 1.29}_{- 1.02}$ & $<6.87$ &   39.57$^{+11.72}_{- 9.23}$ &  -1.00 & QSO  & c \\
J141811.2+523011 & GWS-C50   &  23.96 & 2.910  & 0.33 &   15.60$^{+ 5.44}_{- 4.32}$ & $1.0\times10^{-8}$ &   10.24$^{+ 2.93}_{- 2.33}$ &    2.47$^{+ 0.99}_{- 0.73}$ &    9.70$^{+ 4.44}_{- 3.17}$ & 22.27$^{+ 8.97}_{- 6.60}$ & -0.16 & GAL  & c \\ 
\hline
\end{tabular}
\end{table}
\end{landscape}
%%%%%%%%%%%%%%%%%%%%%%%%%

\label{lastpage}


\begin{thebibliography}{99}

\bibitem[\protect\citeauthoryear{Adelberger \& 
Steidel}{2000}]{adelberger00} Adelberger K.~L., Steidel C.~C., 2000, 
ApJ, 544, 218 
\bibitem[\protect\citeauthoryear{Akiyama et 
al.}{2000}]{akiyama00} Akiyama M., et al., 2000, ApJ, 532,
700 
\bibitem[\protect\citeauthoryear{Alexander et 
al.}{2001}]{alexander01} Alexander D.~M., Brandt W.~N., 
Hornschemeier A.~E., Garmire G.~P., Schneider D.~P., Bauer F.~E., Griffiths 
R.~E., 2001, AJ, 122, 2156 
\bibitem[\protect\citeauthoryear{Alexander et 
al.}{2002}]{alexander02} Alexander D.~M., Aussel H., Bauer F.~E., 
Brandt W.~N., Hornschemeier A.~E., Vignali C., Garmire G.~P., Schneider 
D.~P., 2002, ApJ, 568, L85 
\bibitem[\protect\citeauthoryear{Alexander et al.}{2003}]{A03} Alexander D.~M.~et 
al. 2003, AJ, 126, 539 
\bibitem[\protect\citeauthoryear{Alexander et 
al.}{2005}]{alexander05} Alexander D.~M., Smail I., Bauer F.~E., 
Chapman S.~C., Blain A.~W., Brandt W.~N., Ivison R.~J., 2005, Natur, 434, 
738 
\bibitem[\protect\citeauthoryear{Antonucci \& 
Miller}{1985}]{antonucci85} Antonucci R.~R.~J., Miller J.~S., 1985, 
ApJ, 297, 621 
\bibitem[\protect\citeauthoryear{Awaki et al.}{1991}]{awaki91} 
Awaki H., Koyama K., Inoue H., Halpern J.~P., 1991, PASJ, 43, 195 
\bibitem[\protect\citeauthoryear{Barger et al.}{2001}]{barger01} 
Barger A.~J., Cowie L.~L., Steffen A.~T., Hornschemeier A.~E., Brandt 
W.~N., Garmire G.~P., 2001, ApJ, 560, L23 
\bibitem[\protect\citeauthoryear{Barger et al.}{2002}]{barger02} 
Barger A.~J., Cowie L.~L., Brandt W.~N., Capak P., Garmire G.~P., 
Hornschemeier A.~E., Steffen A.~T., Wehner E.~H., 2002, AJ, 124, 1839 
\bibitem[\protect\citeauthoryear{Barger et al.}{2003}]{barger03} 
Barger A.~J., et al., 2003, AJ, 126, 632 
\bibitem[\protect\citeauthoryear{Basu-Zych \& 
Scharf}{2004}]{basu-zych04} Basu-Zych A., Scharf C., 2004, ApJ, 
615, L85 
\bibitem[\protect\citeauthoryear{Bauer et al.}{2002}]{bauer02} 
Bauer F.~E., Alexander D.~M., Brandt W.~N., Hornschemeier A.~E., Vignali 
C., Garmire G.~P., Schneider D.~P., 2002, AJ, 124, 2351 
\bibitem[\protect\citeauthoryear{Blackburn}{1995}]{blackburn95} 
Blackburn J.~K., 1995, ASPC, 77, 367 
\bibitem[\protect\citeauthoryear{Brandt et al.}{2001}]{brandt01} Brandt W.~N., 
Hornschemeier  A.~E., Schneider  D.~P., Alexander  D.~M., Bauer  F.~E., 
Garmire  G.~P., Vignali  C., 2001, ApJL, 558, L5 
\bibitem[\protect\citeauthoryear{Brusa et al.}{2005}]{brusa05} 
Brusa M., et al., 2005, A\&A, 432, 69
\bibitem[\protect\citeauthoryear{Calzetti}{1997}]{calzetti97} 
Calzetti D., 1997, AJ, 113, 162 
\bibitem[\protect\citeauthoryear{Cash}{1979}]{cash79} Cash W., 
1979, ApJ, 228, 939 
\bibitem[\protect\citeauthoryear{Chapman et 
al.}{2003}]{chapman03} Chapman S.~C., Blain A.~W., Ivison R.~J., 
Smail I.~R., 2003, Natur, 422, 695 
\bibitem[\protect\citeauthoryear{Cowie et al.}{2002}]{cowie02} 
Cowie L.~L., Garmire G.~P., Bautz M.~W., Barger A.~J., Brandt W.~N., 
Hornschemeier A.~E., 2002, ApJ, 566, L5 
\bibitem[\protect\citeauthoryear{David, Jones, \& 
Forman}{1992}]{david92} David L.~P., Jones C., Forman W., 1992, 
ApJ, 388, 82 
\bibitem[\protect\citeauthoryear{Dickey \& 
Lockman}{1990}]{dickey90} Dickey J.~M., Lockman F.~J., 1990, 
ARA\&A, 28, 215
\bibitem[\protect\citeauthoryear{Fabian et al.}{2000}]{fabian00} 
Fabian A.~C., et al., 2000, MNRAS, 315, L8 
\bibitem[\protect\citeauthoryear{Farrah et al.}{2002}]{farrah02} 
Farrah D., Serjeant S., Efstathiou A., Rowan-Robinson M., Verma A., 2002, 
MNRAS, 335, 1163 
\bibitem[\protect\citeauthoryear{Ferrarese \& 
Merritt}{2000}]{ferrarese00} Ferrarese L., Merritt D., 2000, ApJ, 
539, L9 
\bibitem[\protect\citeauthoryear{Gebhardt et 
al.}{2000}]{gebhardt00} Gebhardt K., et al., 2000, ApJ, 539, L13 
\bibitem[\protect\citeauthoryear{Gehrels}{1986}]{gehrels86} 
Gehrels N.\ 1986, ApJ, 303,  336 
\bibitem[\protect\citeauthoryear{George et al.}{2000}]{george00} 
George I.~M., Turner T.~J., Yaqoob T., Netzer H., Laor A., Mushotzky R.~F., 
Nandra K., Takahashi T., 2000, ApJ, 531, 52 
\bibitem[\protect\citeauthoryear{Giacconi et 
al.}{2001}]{giacconi01} Giacconi R., et al., 2001, ApJ, 551, 624 
\bibitem[\protect\citeauthoryear{Giavalisco}{2002}]{giavalisco02}
Giavalisco M., 2002, ARA\&A, 40, 579 
\bibitem[\protect\citeauthoryear{Grimm, Gilfanov, \& 
Sunyaev}{2003}]{grimm03} Grimm H.-J., Gilfanov M., Sunyaev R., 
2003, MNRAS, 339, 793 
\bibitem[\protect\citeauthoryear{Hopkins et 
al.}{2005}]{hopkins05} Hopkins P.~F., Hernquist L., Cox T.~J., Di 
Matteo T., Martini P., Robertson B., Springel V., 2005, ApJ, 630, 705 
\bibitem[\protect\citeauthoryear{Hornschemeier et 
al.}{2001}]{horn01} Hornschemeier A.~E., et al., 2001, ApJ, 
554, 742 
\bibitem[\protect\citeauthoryear{Hornschemeier et 
al.}{2003}]{horn03} Hornschemeier A.~E., et al., 2003, AJ, 
126, 57 5
\bibitem[\protect\citeauthoryear{Huang et al.}{2005}]{huang05} 
Huang J.-S., et al., 2005, ApJ, 634, 137 
\bibitem[\protect\citeauthoryear{Hunt et al.}{2004}]{hunt04} 
Hunt M.~P., Steidel C.~C., Adelberger K.~L., Shapley A.~E., 2004, ApJ, 605, 
625 
\bibitem[\protect\citeauthoryear{Ivison et al.}{2000}]{ivison00} 
Ivison R.~J., Smail I., Barger A.~J., Kneib J.-P., Blain A.~W., Owen F.~N., 
Kerr T.~H., Cowie L.~L., 2000, MNRAS, 315, 209 
\bibitem[\protect\citeauthoryear{Ivison et al.}{2002}]{ivison02} 
Ivison R.~J., et al., 2002, MNRAS, 337, 1 
\bibitem[\protect\citeauthoryear{Koekemoer et 
al.}{2004}]{koekemoer04} Koekemoer A.~M., et al., 2004, ApJ, 600, 
L123 
\bibitem[\protect\citeauthoryear{Laird et al.}{2005}]{L05} 
Laird E.~S., Nandra K., Adelberger K.~L., Steidel C.~C., Reddy N.~A., 2005, 
MNRAS, 359, 47 
\bibitem[\protect\citeauthoryear{Lehmann et 
al.}{2001}]{lehmann01} Lehmann I., et al., 2001, A\&A, 371, 833 
\bibitem[\protect\citeauthoryear{Lehmer et al.}{2005}]{lehmer05} 
Lehmer B.~D., et al., 2005, AJ, 129, 1 
\bibitem[\protect\citeauthoryear{Lowenthal et 
al.}{1997}]{lowenthal97} Lowenthal J.~D., et al., 1997, ApJ, 481, 
673 
\bibitem[\protect\citeauthoryear{Maccacaro et 
al.}{1988}]{macc88} Maccacaro T., Gioia I.~M., Wolter A., 
Zamorani G., Stocke J.~T., 1988, ApJ, 326, 680 
\bibitem[\protect\citeauthoryear{Mainieri et 
al.}{2002}]{mainieri02} Mainieri V., Bergeron J., Hasinger G., 
Lehmann I., Rosati P., Schmidt M., Szokoly G., Della Ceca R., 2002, A\&A, 
393, 425  
\bibitem[\protect\citeauthoryear{Maiolino \& 
Rieke}{1995}]{maiolino95} Maiolino R., Rieke G.~H., 1995, ApJ, 
454, 95 
\bibitem[\protect\citeauthoryear{Magorrian et 
al.}{1998}]{magorrian98} Magorrian J., et al., 1998, AJ, 115, 2285 
\bibitem[\protect\citeauthoryear{Marshall et al.}{2004}]{marshall04}
Marshall H.~L., Tennant A., Grant C.~E., Hitchcock A.~P., O'Dell S.~L., 
Plucinsky P.~P., 2004, SPIE, 5165, 497 
\bibitem[\protect\citeauthoryear{Moran, Filippenko, \& 
Chornock}{2002}]{moran02} Moran E.~C., Filippenko A.~V., 
Chornock R., 2002, ApJ, 579, L71 
\bibitem[\protect\citeauthoryear{Nandra \& 
Pounds}{1994}]{nandra94} Nandra K., Pounds K.~A., 1994, MNRAS, 
268, 405 
\bibitem[\protect\citeauthoryear{Nandra, Laird, \& 
Steidel}{2005}]{nandra05a} Nandra K., Laird E.~S., Steidel C.~C., 
2005, MNRAS, 360, L39 
\bibitem[\protect\citeauthoryear{Nandra et al.}{2005}]{nandra05b} 
Nandra K., et al., 2005, MNRAS, 356, 568 
\bibitem[\protect\citeauthoryear{Nandra et al.}{2002}]{N02} Nandra
K., Mushotzky  R.~F., Arnaud  K., Steidel  C.~C., Adelberger  K.~L.,
Gardner  J.~P., Teplitz  H.~I., Windhorst  R.~A., 2002, ApJ, 576, 625
\bibitem[\protect\citeauthoryear{Page et al.}{2001}]{page01} 
Page M.~J., Stevens J.~A., Mittaz J.~P.~D., Carrera F.~J., 2001, Sci, 294, 
2516 
\bibitem[\protect\citeauthoryear{Page et al.}{2004}]{page04} 
Page M.~J., Stevens J.~A., Ivison R.~J., Carrera F.~J., 2004, ApJ, 611, L85 
\bibitem[\protect\citeauthoryear{Peterson et 
al.}{2006}]{peterson06} Peterson K.~C., Gallagher S.~C., 
Hornschemeier A.~E., Muno M.~P., Bullard E.~C., 2006, AJ, 131, 133 
\bibitem[\protect\citeauthoryear{Pettini et 
al.}{1998}]{pettini98} Pettini M., Kellogg M., Steidel C.~C., 
Dickinson M., Adelberger K.~L., Giavalisco M., 1998, ApJ, 508, 539 
\bibitem[\protect\citeauthoryear{Ptak et al.}{1999}]{ptak99} 
Ptak A., Serlemitsos P., Yaqoob T., Mushotzky R., 1999, ApJS, 120, 179 
\bibitem[\protect\citeauthoryear{Ranalli et al.}{2003}]{ranalli03}
Ranalli  P., Comastri  A., Setti  G., 2003, A\&A, 399, 39 
\bibitem[\protect\citeauthoryear{Reddy \& 
Steidel}{2004}]{reddy04} Reddy N.~A., Steidel C.~C., 2004, ApJ, 
603, L13
\bibitem[\protect\citeauthoryear{Reddy et al.}{2006}]{reddy06} 
Reddy N.~A., Steidel C.~C., Fadda D., Yan L., Pettini M., Shapley A.~E., 
Erb D.~K., Adelberger K.~L., 2006, accepted for publication in ApJ 
\bibitem[\protect\citeauthoryear{Reddy et al.}{2005}]{reddy05} 
Reddy N.~A., Erb D.~K., Steidel C.~C., Shapley A.~E., Adelberger K.~L., 
Pettini M., 2005, ApJ, 633, 748
\bibitem[\protect\citeauthoryear{Rosati et al.}{2002}]{rosati02} 
Rosati P., et al., 2002, ApJ, 566, 667 
\bibitem[\protect\citeauthoryear{Rowan-Robinson}{2000}]{rowan00} Rowan-Robinson M., 2000, MNRAS, 316, 885
\bibitem[\protect\citeauthoryear{Sanders \& 
Mirabel}{1996}]{sanders96} Sanders D.~B., Mirabel I.~F., 1996, 
ARA\&A, 34, 749 
\bibitem[\protect\citeauthoryear{Schmidt et 
al.}{1998}]{schmidt98} Schmidt M., et al., 1998, A\&A, 329, 495 
\bibitem[\protect\citeauthoryear{Seibert, Heckman, \& 
Meurer}{2002}]{seibert02} Seibert M., Heckman T.~M., Meurer 
G.~R., 2002, AJ, 124, 46 
\bibitem[\protect\citeauthoryear{Smail et al.}{2002}]{smail02} 
Smail I., Ivison R.~J., Blain A.~W., Kneib J.-P., 2002, MNRAS, 331, 495 
\bibitem[\protect\citeauthoryear{Stark et al.}{1992}]{stark92} Stark
A.~A., Gammie C.~F., Wilson  R.~W., Bally  J., Linke  R.~A., Heiles
C., Hurwitz  M., 1992, ApJS, 79, 77 
\bibitem[\protect\citeauthoryear{Steidel \& 
Hamilton}{1993}]{steidel93} Steidel C.~C., Hamilton D., 1993, AJ, 
105, 2017 
\bibitem[\protect\citeauthoryear{Steidel et 
al.}{1996}]{steidel96} Steidel C.~C., Giavalisco M., Pettini M., 
Dickinson M., Adelberger K.~L., 1996, ApJ, 462, L17 
\bibitem[\protect\citeauthoryear{Steidel et 
al.}{1999}]{steidel99} Steidel C.~C., Adelberger K.~L., 
Giavalisco M., Dickinson M., Pettini M., 1999, ApJ, 519, 1 
\bibitem[\protect\citeauthoryear{Steidel et 
al.}{2000}]{steidel00} Steidel C.~C., Adelberger K.~L., Shapley 
A.~E., Pettini M., Dickinson M., Giavalisco M., 2000, ApJ, 532, 170 
\bibitem[\protect\citeauthoryear{Steidel et 
al.}{2002}]{steidel02} Steidel C.~C., Hunt M.~P., Shapley A.~E., 
Adelberger K.~L., Pettini M., Dickinson M., Giavalisco M., 2002, ApJ, 576, 
653
\bibitem[\protect\citeauthoryear{Steidel et al.}{2003}]{steidel03} Steidel  C.~C., 
Adelberger  K.~L., Shapley  A.~E., Pettini  M., Dickinson  M.,
Giavalisco  M., 2003, ApJ, 592, 728
\bibitem[\protect\citeauthoryear{Steidel et 
al.}{2004}]{steidel04} Steidel C.~C., Shapley A.~E., Pettini M., 
Adelberger K.~L., Erb D.~K., Reddy N.~A., Hunt M.~P., 2004, ApJ, 604, 534 
\bibitem[\protect\citeauthoryear{Stocke et al.}{1991}]{stocke91} 
Stocke J.~T., Morris S.~L., Gioia I.~M., Maccacaro T., Schild R., Wolter 
A., Fleming T.~A., Henry J.~P., 1991, ApJS, 76, 813 
\bibitem[\protect\citeauthoryear{Veilleux et 
al.}{1995}]{veilleux95} Veilleux S., Kim D.-C., Sanders D.~B., 
Mazzarella J.~M., Soifer B.~T., 1995, ApJS, 98, 171 
\bibitem[\protect\citeauthoryear{Vijh, Witt, \& 
Gordon}{2003}]{vijh03}Vijh U.~P., Witt A.~N., Gordon K.~D., 
2003, ApJ, 587, 533 
\end{thebibliography}
\end{document}